\documentclass[aps,nofootinbib,twocolumn,floatfix,prd,superscriptaddress,letterpaper]{revtex4} 
\usepackage{graphicx,amsmath,amssymb,color,bm,hyperref}
\def\be{\begin{equation}}
\def\ee{\end{equation}}
\def\bea{\begin{eqnarray}}
\def\eea{\end{eqnarray}}
\newcommand{\vs}{\nonumber\\}
\def\ba#1\ea{\begin{align}#1\end{align}}

\newcommand{\g}{\gamma}
\newcommand{\refeq}[1]{Eq.~(\ref{eq:#1})}          
\newcommand{\refeqs}[2]{Eqs.~(\ref{eq:#1})--(\ref{eq:#2})}          
          
\newcommand{\reffig}[1]{Fig.~\ref{fig:#1}}          

\newcommand{\refsec}[1]{\S~\ref{sec:#1}}          
\newcommand{\refapp}[1]{App.~\ref{app:#1}}

\renewcommand{\v}[1]{\mathbf{#1}}

\newcommand{\vx}{\v{x}}
\newcommand{\vk}{\v{k}}

\newcommand{\<}{\langle}
\renewcommand{\>}{\rangle}

\renewcommand{\k}{\kappa}
\renewcommand{\d}{\delta}
\newcommand{\D}{\Delta}

\newcommand{\nhat}{\hat{n}}
\newcommand{\vnhat}{\v{\hat{n}}}

\newcommand{\zt}{\tilde{z}}

\newcommand{\chib}{\bar{\chi}}
\newcommand{\chit}{\tilde{\chi}}

\newcommand{\iMpch}{\:h/{\rm Mpc}}

\def\M{\mathcal{M}}
\def\P{\mathcal{P}}

\def\O{\mathcal{O}}

\def\A{\mathcal{A}}
\def\B{\mathcal{B}}
\def\C{\mathcal{C}}

\def\Del{\eth}
\renewcommand{\Re}{{\rm Re}\,}
\renewcommand{\Im}{{\rm Im}\,}

\def\nhat{\hat{n}}
\def\vnhat{\hat{\v{n}}}

\begin{document}

\title{Cosmic Rulers}

\author{Fabian Schmidt}
\email{fabians@caltech.edu}
\affiliation{Department of Astrophysical Sciences, Princeton University, Princeton, NJ~08540, USA}
\affiliation{Theoretical Astrophysics,
	California Institute of Technology, 
	Mail Code 350-17, Pasadena, CA  91125, USA}
\author{Donghui Jeong}
\email{djeong@pha.jhu.edu}
\affiliation{Department of Physics and Astronomy, Johns Hopkins University, 3400 N. Charles St., Baltimore, MD 21210, USA}

\begin{abstract}
 We derive general covariant expressions for the six
 independent observable modes of distortion of ideal standard rulers in a 
 perturbed Friedmann-Robertson-Walker spacetime.  
 Our expressions are gauge-invariant and valid on the full sky.
 These six modes are most naturally classified in terms of 
 their rotational properties on the sphere, yielding two scalars, 
 two vector (spin-1), and two tensor (spin-2) components.  
 One scalar corresponds to the magnification, while the spin-2
 components correspond to the shear. The vector components allow for a 
 polar/axial decomposition analogous to the $E/B$-decomposition for the 
 shear.  Scalar modes do not contribute to the axial ($B$-)vector, opening 
 a new avenue to probing tensor modes. 
 Our results apply, but are not limited to, the distortion of
 correlation functions (of the CMB, 21cm emission, or galaxies) 
 as well as to weak lensing shear and magnification, 
 all of which can be seen as methods relying on ``standard rulers''.  
\end{abstract}

\date{\today}


\maketitle

\section{Introduction}
\label{sec:intro}

One of the primary goals of cosmology is to accurately measure
the expansion history and the growth of structure in the Universe.  
Many of the cosmological probes used for this purpose can be
classified as standard candles or standard rulers.  
The most obvious examples are
type Ia Supernovae and the baryon acoustic oscillation feature in galaxy
correlation functions, which in an unperturbed Friedmann-Robertson-Walker
(FRW) Universe directly measure the geometry
and expansion history of the Universe \cite{RiessEtal,PerlmutterEtal,EisensteinEtal,2dF}.  Beyond the background cosmology,
cosmological perturbations affect the apparent scale of rulers, which can 
be used as a probe of structure in the Universe.  In fact, from this
point of view, standard rulers
comprise a much larger set of observations: for example, galaxy redshift
surveys measure the correlation function of galaxies, which
is then compared with predictions based on a cosmological model; in other
words, the correlation length of galaxies (or any characteristic scale in their
correlation function) serves as a standard ruler.  
Weak lensing shear, measured using galaxy ellipticities, uses the fact
that galaxies sizes measured along fixed directions are on average equal.  
On the other hand, lensing magnification measurements rely on the fact
that galaxies have a characteristic luminosity (standard candle) and/or
size (standard ruler).  Of course, in the latter three cases the ``ruler'' has
a large amount of scatter, so that one might call it a ``statistical ruler''.  
Another example of this kind is lensing reconstruction on
diffuse backgrounds such as the cosmic microwave background (CMB) 
or 21cm emission from the dark ages \cite{LewisChallinor06,ZahnZaldarriaga,LuPen}.  
In this approach one uses the 
intrinsic correlation pattern of the background, which is known statistically, 
to reconstruct the distortion from the observed pattern.  

There is a simple, unified description of these various cosmological 
probes:  we observe photons from two different directions and redshifts,
which correspond to a known physical scale (e.g., the comoving sound horizon
at recombination, or the characteristic size of a galaxy).  In this paper,
we study in a general covariant setting which underlying properties of the 
spacetime can be measured with an ideal standard ruler, working to linear
order in perturbations.  Since we 
have six parameters to vary when scanning over photon arrival directions
and redshifts, we can measure six degrees of freedom.  These can be
interpreted as the components of a metric (of Euclidean signature) 
mapping apparent coordinate 
distances into actual physical separations at the source.  
It is useful to further decompose these components into parts 
parallel to the line of sight (longitudinal), transverse, and 
mixed longitudinal-transverse parts.  This is equivalent to a decomposition
into scalars, vectors, and tensors in the \emph{two-dimensional} subspace
perpendicular to the photon 4-momentum and the observer's four-velocity; i.e. we
classify components
in terms of their transformation properties under a rotation around the
line of sight.  Note that this is independent of the usual
decomposition of metric perturbations on \emph{three-dimensional}
spatial hypersurfaces (i.e., in terms of the transformation of
plane-wave metric perturbations under a rotation around the $k$-vector).  
We will denote the latter (``3-scalars'' and so on) as $C,\,B_i,\,A_{ij},...$, 
and the former (``2-scalars'' etc) as $\C,\,\B_i,\,\A_{ij}, ...$.    
The transverse components of the distortion, $\A_{ij}$, are
perhaps best known.  They correspond to the magnification (2-scalar) and
shear (2-tensor).  
We show that in general one can also measure a longitudinal 
scalar and the two components of a vector on the sphere.  

On scales much smaller than the horizon, an effective Newtonian
description is sufficient, and this is what essentially all previous
studies are based on.  However,
upcoming surveys will probe scales approaching
the horizon, and an interpretation of these data sets can in principle 
be hampered by gauge ambiguities.  In the case of 
the correlation of galaxy density contrast,
this issue has attracted significant interest and has
recently been resolved
\cite{yoo/etal:2009,challinor/lewis:2011,bonvin/durrer:2011,gaugePk}.  
The unified treatment presented here resolves
these issues for the wide set of cosmological observables mentioned above.  
More precisely, we obtain general
coordinate-independent and gauge-invariant results for all observables,
including the shear and magnification.  
The differential equation (\emph{optical equation})
governing the magnification and shear was first derived in \cite{sachs:1961}.  
The magnification has been derived to first order in \cite{sasaki:1987}.  
The shear has been derived to second order in conformal-Newtonian gauge
in \cite{BernardeauBonvinVernizzi}, while \cite{PitrouEtal} derive the shear for 
general backgrounds.    To the best of our knowledge,
the expression for the observable shear
written in a general gauge is presented here for the first time.  
Further, all expressions are valid on the full sky.
Our approach naturally includes the ``metric shear'' contribution
introduced in \cite{DodelsonEtal}, and we provide a straightforward physical
interpretation of our result.  

In addition, we show how the (2-)vector observable uncovered here
can be decomposed into $E$- and $B$-modes in analogy with the shear,
corresponding to polar and axial vector parts.  As in the case of shear
and CMB polarization, (3-)scalar perturbations do not contribute to the $B$-mode,
while (3-)tensor perturbations contribute.  This in principle offers another
avenue to search for a stochastic gravitational wave background in
large-scale structure, since no scalar perturbations contribute at linear
order.  However, a spectroscopic data set is likely
necessary to reconstruct the vector component with an interesting 
signal-to-noise.  

Apart from the linear treatment of metric perturbations, we
make two further simplifying assumptions:  first, we assume ``small rulers''
in the sense that rulers subtend a small apparent angle and redshift interval.  
Wide-angle effects are likely negligible for almost all applications (the
large-scale BAO feature being perhaps the most important exception).  
A treatment of wide-angle effects necessarily involves a detailed model
of the survey geometry, which is clearly beyond the scope of this paper.  
The second assumption is that any scatter or variation in the actual 
(``intrinsic'') physical scale of the standard ruler is uncorrelated
with large-scale perturbations.  This will not hold true in general,
since the physical systems used as rulers will be affected by their
large-scale environment.  One well-known example is the intrinsic
alignment contribution to shear correlations \cite{CatelanKamionkowskiBlandford}.  
Further examples include
the distortion of correlation
functions by large-scale tidal fields \cite{PenEtal}, 
or by a non-Gaussian coupling of the density field to primordial 
degrees of freedom \cite{jeong/kamionkowski:2012}.  
Since these ``intrinsic effects'' depend on the physics of the given ruler, 
we refrain from discussing them here, as they would distract from the
generality of the rest of the results.  

Finally, while we focus on general standard rulers here, the case of
standard candles is directly related to our results.  This is because
the relation between angular diameter distance $D_A$ and luminosity
distance $D_L$,
\be
D_L = (1+\zt)^2 D_A,
\label{eq:DLDA}
\ee
where $\zt$ is the observed redshift, holds in a general spacetime
and for any source (this is a consequence of photon phasespace conservation).  
Thus, the magnification measured for standard candles is identical
to the magnification for standard rulers which we will derive here.  
However, as discussed, standard rulers can 
measure five additional degrees of freedom not accessible to standard candles.  

Our results are of immediate relevance to recent studies which consider
the $B$-modes of the cosmic shear as possible probe of an inflationary
gravitational wave background \cite{DodelsonEtal,Dodelson10,paperII},
and to studies that propose to use the high-redshift 21cm emission 
for the same purpose \cite{MasuiPen,BookMKFS}.  In particular,
our expressions can be used directly to construct optimal estimators 
on the full sky for
searching for the imprint of gravitational waves in a three-dimensional
field such as the 21cm background.  
We also use many of the results derived here in two recent papers studying 
the impact of gravitational waves on the observed large-scale structure
\cite{paperI,paperII}.  

The outline of the paper is as follows: we begin in \refsec{not} by
introducing our metric convention and useful notation.  The general expression 
for the mapping from apparent size to true physical size of the ruler
is derived in \refsec{ruler}.  We then decompose the contributions
into longitudinal and transverse parts in \refsec{SVT}.  The following 
three sections deal with these different parts consecutively.  
We discuss and conclude in \refsec{disc}.  The appendix contains
a large amount of additional reference material on multipole expansions of
higher spin functions, perturbed photon geodesic equation, and various
test cases applied to our results.  

\section{Notation}
\label{sec:not}

In a general gauge, the perturbed FRW metric is given by
\ba
ds^2 = a^2(\eta)\Big[&
-(1+2A) d\eta^2 - 2B_i d\eta dx^i 
\vs
& + \left(\d_{ij}+h_{ij}\right) dx^i dx^j \Big],
\label{eq:metric}
\ea
where we have assumed a spatially flat Universe (curvature can be
included straightforwardly, at the expense of some extra notation).  
Here, $\eta$ denotes conformal time.  
Often, the spatial part is further expanded as
\be
h_{ij} = 2D \d_{ij} + 2 E_{ij},
\label{eq:DE}
\ee
where $E_{ij}$ is traceless.  
We shall also present the most interesting results in two popular
gauges: the synchronous-comoving (sc) gauge, where $A = 0 = B_i$, so that
\be
ds^2 = a^2(\eta)\left[- d\eta^2
+ \left(\d_{ij}+h_{ij}\right) dx^i dx^j \right];
\label{eq:metric_sc}
\ee
and the conformal-Newtonian (cN) gauge, where $B_i = 0 = E_{ij}$.  In the
latter case, we  denote $A = \Psi$, $D = \Phi$, conforming 
with standard notation, so that
\be
ds^2 = a^2(\eta)\left[- (1+2\Psi) d\eta^2
+ (1+2\Phi) \d_{ij} dx^i dx^j \right].
\label{eq:metric_cN}
\ee
We also denote the background FRW metric (in the absence of perturbations) as 
$\bar g_{\mu\nu} = a^2(\eta) \eta_{\mu\nu}$.  

It is useful to define projection operators parallel and perpendicular
to the observed line-of-sight direction $\nhat^i$,
so that for any spatial vector $X^i$ and tensor $E_{ij}$,
\ba
X_\parallel \equiv\:& \nhat_i X^i, \vs
E_\parallel \equiv\:& \nhat_i \nhat_j E^{ij}, \vs
X_\perp^i \equiv\:& \P^{ij} X_j \vs
\P^{ij} \equiv\:& \d^{ij} - \nhat^i \nhat^j.
\label{eq:proj1}
\ea
Correspondingly, we define projected derivative operators,
\ba
\partial_\parallel \equiv\:& \nhat^i\partial_i,{\rm ~and} \vs 
\partial_\perp^i \equiv\:& \P^{ij} \partial_j.
\label{eq:proj2}
\ea
Note that $\partial_\perp^i,\,\partial_\parallel$ and $\partial_\perp^i,\,\partial_\perp^j$
do not commute.  Further, we find
\be
\partial_j \nhat^i = \partial_{\perp j} \nhat^i = \frac1\chi \P_j^{\  i},
\ee
where $\chi$ is the norm of the position vector so that $\nhat^i = x^i/\chi$.  
Note that $\nhat^i$ and $\partial_\parallel$ commute.  
More expressions can be found in \S~II of \cite{gaugePk}.

Finally, it proves useful to decompose the quantities defined on the
sphere, i.e. as function of the unit line-of-sight vector $\vnhat$, 
in terms of their properties under a rotation around $\vnhat$.  In particular,
consider an orthonormal coordinate system $(\v{e}_1,\v{e}_2,\vnhat)$.  
If we rotate the coordinate system around $\vnhat$ by an angle $\psi$,
so that $\v{e}_i \to \v{e}'_i$,
then the linear combinations $\v{m}_\pm \equiv (\v{e}_1\mp i\,\v{e}_2)/\sqrt{2}$
transform as 
\be
\v{m}_\pm \to \v{m}'_\pm = e^{\pm i\psi} \v{m}_\pm.
\label{eq:spin1}
\ee
We say that a general function $f(\vnhat)$ is spin-$s$ if it transforms
under the same transformation as
\be
f(\vnhat) \to f(\vnhat)' = e^{i s\psi} f(\vnhat).
\ee
An ordinary scalar function on the sphere is clearly spin-0, while the
unit vectors $\v{m}_\pm$ defined above are spin$\pm1$ fields.  More details
can be found in \refapp{spinSH}.  This decomposition is particularly
useful for deriving multipole coefficients and angular power spectra.  
We also define
\ba
X_\pm \equiv\:& m_\mp^i X_i \vs
E_\pm \equiv\:& m_\mp^i m_\mp^j E_{ij}
\label{eq:Xpm}
\ea
for any 3-vector $X_i$ and 3-tensor $E_{ij}$.  

For the quantitative results shown in \reffig{Cl}, we assume
a flat $\Lambda$CDM cosmology with $h=0.72$, $\Omega_m=0.28$, a scalar
spectral index $n_s=0.958$ and power spectrum normalization at $z=0$ of
$\sigma_8 = 0.8$.

\section{Standard ruler}
\label{sec:ruler}

\begin{figure}[t!]
\centering
\includegraphics[width=0.45\textwidth]{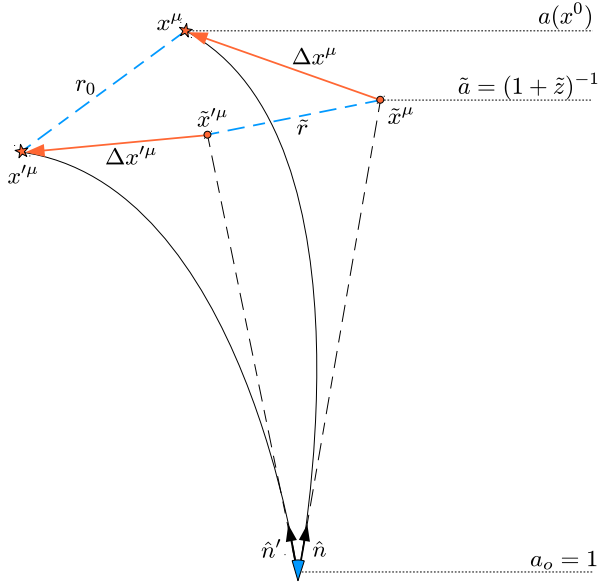}
\caption{Illustration of the apparent and actual standard ruler.  Photons
arrive out of the observed directions $\vnhat,\,\vnhat'$ and with
observed redshifts $\zt,\,\zt'$.  The apparent positions are indicated
by $\tilde x^\mu,\,\tilde x'^\mu$, while the true positions are at
$x^\mu,\,x'^\mu$, perturbed by the displacements $\D x^\mu,\,\D x'^\mu$
(whose magnitude is greatly exaggerated here).  
$\tilde r$ is the apparent size of the ruler, while $r_0$ is the true
ruler.
\label{fig:sketch}}
\end{figure}

In the absence of perturbations, photon geodesics are given by straight lines
in conformal coordinates,
\be
\bar{x}^\mu(\chi)
=
\left(
\eta_0-\chi, \vnhat\chi
\right),
\label{eq:geod_conf}
\ee
where we have chosen the comoving distance $\chi$ as affine parameter.  
Correspondingly, for a photon arriving from a direction $\vnhat$ with
redshift $\zt$, we assign an ``observed'' position of emission $x^\mu$ given by
\ba
\tilde x^0 =\:& \eta_0 - \chit \vs
\tilde x^i =\:& \nhat^i\,\chit \vs
\chit \equiv\:& \chib(\zt),
\label{eq:xtilde}
\ea
where $\chib(\zt)$ denotes the comoving distance-redshift relation in the
background Universe.  Here we have chosen the observer to reside at the 
spatial origin without loss of generality.  The coordinate time of the observer
who is assumed to be comoving is fixed by the condition of a fixed proper time $t_0$ at observation.  On the other hand, the actual spacetime point of emission, denoted
with $x^\mu$, is displaced from the observed positions by 
$\D x^\mu$ (see also \reffig{sketch}),
\be
x^\mu = \tilde x^\mu + \D x^\mu(\vnhat,\zt).
\ee
Further, we will need the scale factor at emission.  It is related
to the inferred emission scale factor $\tilde a \equiv 1/(1+\zt)$ by
\be
\frac{a(x^0(\vnhat,\tilde z))}{\tilde a} = 1 + \D\ln a(\vnhat,\tilde z).
\label{eq:Dlnadef}
\ee
At first order, $\D\ln a = \D z/(1+\tilde z)$, where $\D z$
is the difference between the observed redshift and the redshift that would
be observed in an unperturbed Universe.  Note that the latter quantity
is gauge-dependent.  
We will give explicit expressions for $\D\ln a$ and $\D x^\mu$ in \refsec{SVT}.

The displacements $\D\ln a$, $\D x^\mu$ are not observable (they depend on which
gauge, or frame, the spacetime perturbations are described in).  
However, we will construct them in such a way that for a local source,
i.e. for photons emitted an infinitesimal distance away from the observer, 
the perturbations vanish\footnote{That is, up to an additive constant to $\D\ln a$ and $\D x^0$ which enforces the condition that the observer is at a fixed proper time.}:
\be
\D\ln a,\  \D x^\mu \stackrel{\zt \to 0}{=} 0.
\ee
In order to determine actual observables, we consider the case of a
standard ruler.  A standard ruler exists if we can identify two
spacetime points which are separated by a fixed spacelike distance $r_0$.  
What we observe is the \emph{apparent} size
at which this ruler appears in a given direction $\vnhat$ and redshift
$\tilde z$.  Let $\vnhat,\tilde z$ and $\vnhat',\tilde z'$ denote the observed 
coordinates of the ``end points'' of the ruler, and $\tilde\vx$ and
$\tilde\vx'$ the apparent spatial positions inferred through \refeq{xtilde}.  
The \emph{inferred} physical separation is then given by
\be
\tilde r^2 = \tilde a^2 \d_{ij} (\tilde x^i - \tilde x'^i) (\tilde x^j - \tilde x'^j),
\label{eq:rt1}
\ee
where $\tilde a = 1/(1+\zt)$ is the observationally inferred scale factor
at emission (\reffig{sketch}).  

We now have to carefully consider what the condition of a standard
ruler in cosmology means.  A useful, physically motivated definition is
that it corresponds to a fixed spatial scale as measured by local
observers which are comoving with the cosmic fluid; precisely, 
the spatial part of the four-velocity $u^\mu$ of these observers is given by
\be
v^i = \frac{T^i_{\  0}}{\rho + p}.  
\label{eq:vcom}
\ee  
We are mostly interested in applications to the large-scale structure during
matter domination; in this case, the cosmic fluid is simply matter
(dark matter + baryons), and there is no ambiguity in this definition.  
In synchronous-comoving gauge, \refeq{vcom} yields $v^i = 0$.  
Further, in the following we assume the ruler scale is fixed.  
An evolving ruler is considered in \cite{Tpaper}.

This definition can also be phrased as that the length of the ruler 
is defined on a surface of
constant proper time of comoving observers.  This proper time corresponds
to the ``local age'' of the Universe.  
The separation of the two endpoints
of the ruler, $x^\mu,\,x'^\mu$, projected onto this hypersurface
should thus be equal to the fixed scale $r_0$:
\ba
\left[g_{\mu\nu}(x^\alpha) + u_\mu(x^\alpha) u_\nu(x^\alpha)\right]
 (x^a-x'^a) (x^b - x'^b) = r_0^2,
\label{eq:r01}
\ea
where $g_{\mu\nu} + u_\mu u_\nu$ is the metric projected perpendicular 
to $u_\mu$, the four-velocity of the comoving observers
(note that $u_\mu u^\mu = -1$).  Here and throughout, we
will assume for simplicity that the ruler is ``small'', i.e. 
it subtends a small angle, and
redshift interval ($|\tilde z-\tilde z'| \ll \tilde z$).  This entails
$\d\tilde x^i \d\tilde x_i \ll \chit^2$, and that we can simply 
evaluate the metric and 
four-velocity at either end-point (corrections involve higher powers of
$x^\mu - x'^\mu$).  

The four-velocity of comoving observers, whose spatial components
are fixed by \refeq{vcom}, is given by
\ba
u^\mu =\:& a^{-1} \left(1-A,\, v^i \right) \vs
u_\mu =\:& a \left(-1-A,\, v_i - B_i \right),
\label{eq:umu}
\ea
where we consider $v^i$ to be first order (as the metric perturbations).  
In the following, we will assume sources to be comoving as well, i.e. to 
follow \refeq{umu}.  It is straightforward to generalize the 
treatment to different source velocities.  
Using \refeq{metric} and
\refeq{umu}, we have
\ba
g_{\mu\nu} + u_\mu u_\nu 
=\:& a^2 \left(\begin{array}{cccc}
0 & & -v_i & \\
 & & & \\
- v_i & & \d_{ij} + h_{ij} & \\
 & & & 
\end{array}\right).
\ea
With this, \refeq{r01} yields
\ba
& -2 \tilde a^2 v_i \left\{ \d\tilde x^0 \d\tilde x^i + \d\tilde x^0 [\D x^i -\D x'^i] + \d\tilde x^i [\D x^0 - \D x'^0 ] \right\} \vs
&+ g_{ij}(x^\alpha) \bigg\{ \d\tilde x^i \d\tilde x^j + \d\tilde x^i [ \D x^j - \D x'^j ] 
+ [ \D x^i - \D x'^i ]\d\tilde x^j
\bigg\}\vs
& = r_0^2,
\label{eq:r02}
\ea
where $\D x^\mu = \D x^\mu(\vnhat,\zt)$, $\D x'^\mu = \D x^\mu(\vnhat',\zt')$,
and the components of the \emph{apparent} separation vector are
\be
\d\tilde x^\mu = \tilde x^\mu - \tilde x'^\mu.
\label{eq:dxmu}
\ee  
In order to evaluate the spatial metric $g_{ij}(x^\alpha)$ at the 
location of the ruler, we use \refeq{Dlnadef} to obtain at first order
\be
g_{ij}(x^\alpha) = \tilde a^2\left[\left(1 + 2 \D\ln a \right) \d_{ij} + h_{ij} \right].
\ee
We now again make use of the ``small ruler'' approximation, so that
\be
\D x^i - \D x'^i \simeq \d\tilde x^\alpha \frac{\partial}{\partial\tilde x^\alpha}
\D x^i.
\ee
Like any vector, we can decompose the spatial part of the apparent 
separation $\d\tilde x^i$ into parts parallel and transverse to the line
of sight:
\ba
\d\tilde x_\parallel \equiv\:& \nhat_i \d\tilde x^i \vs
\d\tilde x_\perp^i \equiv\:& \P^i_{\  j} \d\tilde x^j = \d\tilde x^i - \nhat^i \d\tilde x _\parallel.
\ea
In the correlation function literature, $\d\tilde x_\parallel,\,|\d\tilde\vx_\perp|$
are sometimes referred to as $\pi$ and $\sigma$, respectively.  Then,
\ba
\d\tilde x^\alpha \frac{\partial}{\partial\tilde x^\alpha} =\:&
 (\d\tilde x^0 \partial_\eta + \d\tilde x_\parallel \partial_\parallel) 
+ \d\tilde x_\perp^i \partial_{\perp\,i},
\ea
where we have similarly defined $\partial_\parallel = \nhat^i \partial_i$,
$\partial_{\perp\,i} = \P^{\  j}_i \partial_j$.   Since the observed
coordinates $\tilde x^\mu$ by definition satisfy the light cone condition 
with respect to the unperturbed FRW metric, 
we have $\d\tilde x^0 = -\d\tilde x_\parallel$ in the small-angle approximation.  
Thus,
\ba
\d\tilde x^0 \partial_\eta + \d\tilde x_\parallel \partial_\parallel
=\:& \d\tilde x_\parallel (\partial_\parallel - \partial_\eta) \vs
=\:& \d\tilde x_\parallel \frac{\partial}{\partial\chit} = \d\tilde x_\parallel
H(\zt) \frac{\partial}{\partial\zt},
\ea
where $\partial/\partial\chit$ is the derivative with respect to the affine
parameter at emission.  We thus have
\be
\d\tilde x^\alpha \frac{\partial}{\partial\tilde x^\alpha} = 
\d\tilde x_\parallel \partial_{\chit} + \d\tilde x_\perp^i \partial_{\perp\,i}.
\ee
Working to first order in perturbations, we then obtain 
\ba
r_0^2 - \tilde r^2  =\:& 2 \D\ln a\: \tilde r^2 
+ \tilde a^2  h_{ij} \d\tilde x^i \d\tilde x^j \vs
+& 2 \tilde a^2 \left( v_\parallel  \d\tilde x_\parallel^2 + v_{\perp\,i} \d\tilde x_\perp^i \d\tilde x_\parallel\right)
\vs
+& 2\tilde a^2 \d_{ij} \d\tilde x^{i} \left(\d\tilde x_\parallel \partial_{\chit}
+ \d\tilde x_\perp^k \partial_{\perp\,k} \right) \D x^{j}.
\label{eq:rt2}
\ea
All terms are straightforward to interpret: there are the perturbations
to the metric (both from the metric perturbation $h_{ij}$ and the
perturbation to the scale factor at emission); the contribution $\propto v$
from the projection from fixed-$\eta$ to fixed-proper-time
hypersurfaces; and the difference in
the spatial displacements of the endpoints of the ruler.  

\section{Scalar-vector-tensor decomposition on the sky}
\label{sec:SVT}

\begin{figure}
\centering
\includegraphics[width=0.5\textwidth]{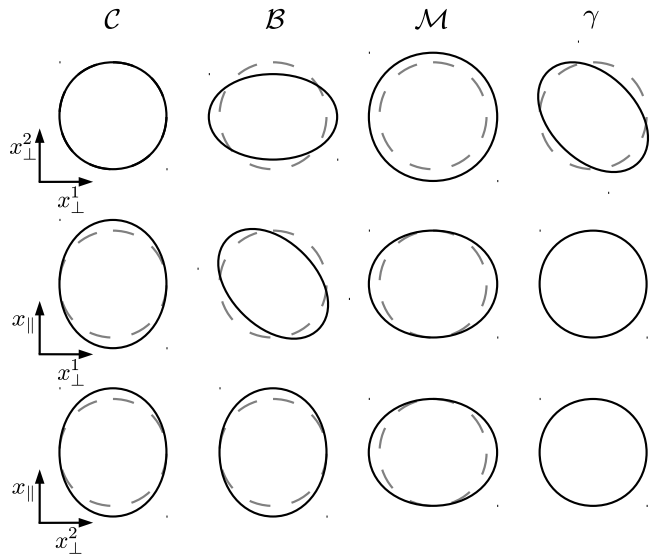}
\caption{Illustration of the distortion of standard rulers
due to the longitudinal (2-)scalar $\mathcal{C}$, (2-)vector $\B$, and 
transverse components, magnification $\M$ and shear $\g$.  
The first row shows the projection onto the sky plane, while the
second (third) row show the projection onto the line-of-sight and
$x_\perp^1$ ($x_\perp^2$) axes, respectively.  In case of $\B$ and $\g$,
we only show one of the two components.  See also Fig.~3 in \cite{sachs:1961}.  
\label{fig:SVT}}
\end{figure}

It is useful to separate the contributions to \refeq{rt2} in terms of the 
observed longitudinal and transverse displacements.  For some
applications, only the transverse displacements are relevant.  This
is the case for diffuse backgrounds without redshift resolution,
such as the CMB or the cosmic infrared background, and largely the case
for photometric galaxy surveys.  On the other hand, spectroscopic
surveys and redshift-resolved backgrounds such as the 21cm emission
from high-redshifts are able to measure the longitudinal displacements
as well.  

Noting that
$\tilde r^2 = \tilde a^2[\d\tilde x_\parallel^2 + (\d\tilde\vx_\perp)^2]$, 
and taking the square root of \refeq{rt2}, we obtain the 
relative perturbation to the physical scale of the ruler as
\ba
\frac{\tilde r - r_0 }{\tilde r} =\:&
\C \frac{(\d\tilde x_\parallel)^2}{\tilde r_c^2} + \B_i \frac{\d\tilde x_\parallel \d\tilde x_\perp^i}{\tilde r_c^2}
+ \A_{ij} \frac{\d\tilde x_\perp^i \d\tilde x_\perp^j}{\tilde r_c^2},
\label{eq:r03}
\ea
where we have defined
$\tilde r_c \equiv \tilde r/\tilde a$ as the apparent comoving size of the ruler. 
The quantities multiplying $\C,\,\B_i,\,\A_{ij}$ are thus simply geometric
factors.  The coefficients are given by
\ba
\C =\:&  - \D\ln a 
- \frac12 h_\parallel - v_\parallel - \partial_{\chit} \D x_\parallel
\vs
\B_i =\:& -\P_i^{\  j} h_{jk} \nhat^k - v_{\perp i} - \nhat^k \partial_{\perp\,i} \D x_k - \partial_{\chit} \D x_{\perp i}
\vs
\A_{ij} =\:& 
- \D\ln a\: \P_{ij}
- \frac12 \P_i^{\  k}\P_j^{\  l} h_{kl} \vs
& - \frac12 \left(\P_{jk} \partial_{\perp\,i} + \P_{ik} \partial_{\perp\,j}\right) \D x^k, \label{eq:coeff}
\ea
where $\D x_\parallel,\,\D x_\perp^i$ are the parallel and
perpendicular components of the displacements $\D x^i$.  
Note that while we
have assumed that the ruler is small (i.e. $\d\tilde x^i \ll \chit$), the 
expressions for $\C,\,\B_i,\,\A_{ij}$ are valid on the 
full sky.  \reffig{SVT} illustrates the distortions induced by these
components.   
Observationally, we have 6 free parameters (assuming accurate redshifts
are available): the
location of one point $\vnhat,\,\zt$, and the separation vector described by
$\d\tilde x^i$ (with $\d\tilde x^0$ being fixed by the light cone condition).  
Using these, we can measure a (2-)scalar on the sphere, $\C$, a $2\times2$ 
symmetric matrix, $\A_{ij}$, and a 2-component vector on the sphere, $\B_i$.  
As a symmetric matrix on the sphere, $\A_{ij}$ has a scalar component, given
by the trace $\M \equiv \P^{ij} \A_{ij}$ (\emph{magnification}), and two 
components of the traceless 
part which transform as spin-2 fields on the sphere (\emph{shear} ${}_{\pm 2}\g$ 
as defined in \refeq{shear1} below).  
These quantities are observable and gauge-invariant, while any of the individual
contributions in \refeq{coeff} are not in general.  Note that we cannot measure
any of the anti-symmetric components, such as the rotation.  This is because
we have not assumed the existence of any preferred directions in the 
Universe.  If there is a primary spin-1 or higher spin field, such as 
the polarization in case of the CMB, then a rotation can be measured as 
it mixes the spin$\pm 2$ components (see, e.g. \cite{gluscevic/etal:2009}).
In the next sections we study these three terms in turn.  

For reference, we now give the explicit expressions for the 
displacements $\D x^i$ and $\D\ln a$.  They are defined such that
$\D x^i = 0 = \D\ln a$ for a local source, i.e. for $\zt \approx 0$, up to a shift in the observer's time coordinate.    
The details of the derivation are presented in \refapp{geod}.  
Separating into line-of-sight and transverse parts, we have
\ba
\D x_\parallel =\:& \int_0^{\chit} d\chi\left[ A - B_\parallel - \frac12 h_\parallel
\right]
- \frac{1+\zt}{H(\zt)} \D \ln a \vs
& -\int_0^{t_0} A(\v{0},t) dt \label{eq:Dxpar}\\
\D x_\perp^i =\:& \left[\frac12 \P^{ij} (h_{jk})_o\, \nhat^k + B^i_{\perp o} - v^i_{\perp o}\right] \chit  \label{eq:Dxperp}\\
& + \int_0^{\chit} d\chi \bigg[
- B_\perp^i - \P^{ij} h_{jk}\nhat^k \vs
&\hspace*{1.6cm} + (\chit-\chi)\bigg\{
- \partial_\perp^i A + \nhat^k \partial_\perp^i B_k \vs
&\hspace*{3.4cm} + \frac12 (\partial_\perp^i h_{jk})\nhat^j\nhat^k
\bigg\}\bigg] \vs
=\:& \left[\frac12 \P^{ij} (h_{jk})_o\, \nhat^k + B^i_{\perp o} - v^i_{\perp o}\right] \chit \label{eq:Dxperp2}\\
& - \int_0^{\chit} d\chi \bigg[
\frac{\chit}{\chi} \left(
B_\perp^i + \P^{ij} h_{jk}\nhat^k\right) 
\vs
&\hspace*{1.6cm}+(\chit-\chi)\partial_\perp^i
\left( A - B_\parallel - \frac12 h_\parallel\right)
\bigg].\nonumber
\ea
The perturbation to the scale factor at emission is given by
\ba
\D\ln a =\:&  A_o - A + v_\parallel - v_{\parallel o} 
+ \int_0^{\chit} d\chi\left[
- A' + \frac12 h_{\parallel}' + B_\parallel'\right] \vs
& - H_0 \int_0^{t_0} A(\v{0},\bar\eta(t)) dt\,.
\label{eq:Dlna}
\ea
Here, a subscript $o$ indicates quantities evaluated at the observer,
while primes denote derivatives with respect to $\eta$.  
Note the appearance of the scalar quantity 
$A - B_\parallel - \frac12 h_\parallel$ in \refeqs{Dxpar}{Dlna}.  
This is the ``lensing potential'' $\Phi-\Psi$ in conformal-Newtonian gauge,
written in the general gauge \refeq{metric}.  The term in the second lines of \refeq{Dxpar} and \refeq{Dlna} comes from requiring the observer to lie at a fixed proper time, rather than at fixed scale factor or coordinate time, which are gauge-dependent quantities.  While these terms only contribute to the monopole of $\C$ and $\M$, which
are typically not observable, they are essential for cross-checking the result with test cases and against gauge-transformations.

In particular, in the two popular gauges
introduced in \refsec{ruler}, \refeq{Dlna} becomes
\ba
(\D\ln a)_{\rm sc} =\:& \frac12 \int_0^{\chit} d\chi\:h_\parallel'
\label{eq:Dlna_sc}\\
(\D\ln a)_{\rm cN} =\:& \Psi_o - \Psi + v_\parallel - v_{\parallel o}
+ \int_0^{\chit} d\chi \left[ \Phi' - \Psi'\right] \vs
& - H_0 \int_0^{t_0} \Psi(\v{0},\bar\eta(t)) dt\,. \label{eq:Dlna_cN}
\ea
The latter result clearly shows the ``Sachs-Wolfe'', ``Doppler'',
and ``integrated Sachs-Wolfe'' contributions, along with the coordinate
time perturbation at the observer fixing the proper time.

\section{Longitudinal scalar}
\label{sec:C}

The longitudinal component can be simplified to become
\ba
\C =\:&  - \D\ln a
\left[1 - H(\zt) \frac{\partial}{\partial\zt}\left( \frac{1+\zt}{H(\zt)}\right)
\right] \vs
& - A - v_\parallel + B_\parallel \vs
& + \frac{1+\zt}{H(\zt)} \left(
 - \partial_\parallel A + \partial_\parallel v_\parallel 
+ B_\parallel' - v_\parallel' + \frac12 h_{\parallel}'  \right).
\label{eq:C}
\ea
The first line contains the contributions due to the fact that the
scale factor at emission is perturbed from $1/(1+\zt)$, and due to the evolution of the distance-redshift relation.  The second
line contains the perturbations from the metric at the source location
($-A$) and the projection from coordinate-time to proper-time hypersurfaces
($B_\parallel-v_\parallel$).  Finally, the contributions from the line-of-sight
derivative of the line-of-sight displacements ($\propto (1+\zt)/H(\zt)$)
are given in the third line.  
Note the term $\partial_\parallel v_\parallel$, which is the dominant term
on small scales in the conformal-Newtonian gauge.  This term is
also responsible for the leading-order redshift distortions \cite{Kaiser87}.  
Apart from the pertubation to the scale factor at emission,
$\C$ does not involve any integral terms; this is expected since $\C$
is the only term remaining if the two lines of sight coincide ($\vnhat=\vnhat'$).  
In this case, the two rays share the same path from the closer of the
two emission points, and no quantities integrated along the line of sight
can contribute to the perturbation of the ruler.  

\begin{figure}[t!]
\centering
\includegraphics[width=0.49\textwidth]{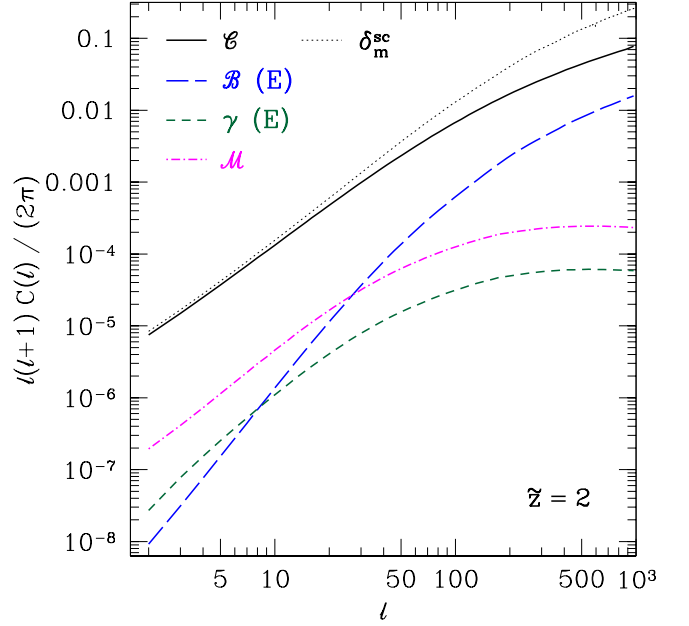}
\caption{Angular power spectra of the different standard ruler perturbations
produced by a standard scale-invariant power spectrum of curvature perturbations:
$\C$, $E$-mode of $\B_i$, $E$-mode of the shear, and magnification $\M$.  
All quantities are calculated for a non-evolving ruler and a sharp source 
redshift of $\zt = 2$.  
For comparison, the thin dotted line shows the angular power spectrum at $z=2$ of the matter 
density field in synchronous-comoving gauge.  Note that all quantities 
shown here, except for $\d_m^{\rm sc}$, are gauge-invariant and (in principle)
observable.
\label{fig:Cl}}
\end{figure}

Restricting to the synchronous-comoving and conformal-Newtonian gauges,
respectively, we obtain
\ba
(\C)_{\rm sc} =\:& - (\D\ln a)_{\rm sc}
\left[1 - H(\zt) \frac{\partial}{\partial\zt}\left( \frac{1+\zt}{H(\zt)}\right)
\right] \vs
& + \frac{1+\zt}{2 H(\zt)}h_{\parallel}'.
\label{eq:C_sc}\\
(\C)_{\rm cN} =\:& - (\D\ln a)_{\rm cN}
\left[1 - H(\zt) \frac{\partial}{\partial\zt}\left( \frac{1+\zt}{H(\zt)}\right)
\right] \vs
& - \Psi - v_\parallel + \frac{1+\zt}{H(\zt)} \left(
 - \partial_\parallel\Psi + \partial_\parallel v_\parallel 
 - v_\parallel' + \Phi'  \right).
\label{eq:C_cN}
\ea
Note that in case of the sc-gauge expression, 
the redshift-space distortion term is included in the last term, through
$h_\parallel'/2 = D' + \partial_\parallel^2 E'$.  \reffig{Cl} shows
the angular power spectrum of $\C$ due to standard adiabatic
scalar perturbations in a $\Lambda$CDM cosmology (the details of the
calculation are given in \refapp{calc}).  Clearly,
$\C$ is of the same order as the matter density contrast in 
synchronous-comoving gauge on all scales.  In particular, 
the velocity gradient term dominates over all other contributions.  
Due to the different dependence on the angle with the line of sight,
the projection kernel of $\C$ is proportional to $\partial_x^2 j_l(x)$,
while that of $\d_m^{\rm sc}$ is $\propto j_l(x)$.  The former favors larger
$x$ at a given $l$, and thus leads to a relative suppression as the
slope of the matter power spectrum changes at $k \gtrsim 0.01 \iMpch$.

\section{Vector}
\label{sec:vector}

Next, we have the two-component vector
\ba
\B_i =\:& -\P_i^{\  j}h_{jk}\nhat^k - v_{\perp i} - \partial_{\perp\,i} \D x_\parallel - \partial_{\chit} \D x_{\perp\,i}
+ \frac{\D x_{\perp\,i}}{\chit}
\vs
=\:& - v_{\perp i} + B_{\perp i} + \frac{1+\zt}{H(\zt)}\partial_{\perp i}\D\ln a,
\label{eq:Bi}
\ea 
where we have inserted projection operators for clarity (these are trivial
since $\B_i$ is contracted with $\d\tilde x_\perp^i$).  
As expected, this vector involves the transverse derivative of the
line-of-sight displacement and the line-of-sight derivative of the
transverse displacement.  Note that these two quantities are 
\emph{not} observable individually.  

Using the spin$\pm1$ unit vectors $\v{m}_\pm$, 
$\B_i$ can be decomposed into spin$\pm1$ components:
\ba
\B_i =\:& {}_{+1}\B m_+^i + {}_{-1}\B m_-^i \vs
{}_{\pm 1}\B \equiv\:& m_\mp^i \B_i 
= - v_{\pm} + B_{\pm} + \frac{1+\zt}{H(\zt)}\partial_{\pm}\D\ln a,
\label{eq:Bpm}
\ea
where we have used the notation of \refeq{Xpm}.  
Similar to before, we can specialize this general result to the
synchronous-comoving and conformal-Newtonian gauges:
\ba
({}_{\pm 1}\B)_{\rm sc} =\:& \frac{1+\zt}{2 H(\zt)}\int_0^{\chit}d\chi\frac{\chi}{\chit}\partial_\pm h_\parallel'
\label{eq:Bpm_sc}\\
({}_{\pm 1}\B)_{\rm cN} =\:& - v_\pm + \frac{1+\zt}{H(\zt)}\partial_{\pm}\D\ln a \vs
=\:& -v_\pm + \frac{1+\zt}{H(\zt)}\Bigg(-\partial_\pm \Psi + \partial_\pm 
[v_\parallel - v_{\parallel o}] \vs
& \hspace*{2.5cm} + \int_0^{\chit} d\chi\frac{\chi}{\chit} \partial_\pm(\Phi'-\Psi')
\Bigg).
\label{eq:Bpm_cN}
\ea
On small scales, the dominant contribution to $\B_i$ comes from the 
transverse derivative of the line-of-sight component of the velocity
$\partial_\pm v_\parallel$, which
is of the same order as the tidal field.  

Applying the spin-lowering operator $\bar\Del$ to ${}_1\B$ (see \refapp{spinSH}) yields
a spin-zero quantity, which can be expanded in terms of the usual
spherical harmonics\footnote{This is of course equivalent to expanding
${}_1\B$ in terms of spin-1 spherical harmonics.}.  We then obtain
the multipole coefficients of $\B$ as
\ba
a^{\B}_{lm}(\zt) =\:& - \sqrt{\frac{(l-1)!}{(l+1)!}} \int d\Omega\: \left[\bar\Del\,{}_1\B(\vnhat,\zt)\right] Y^*_{lm}(\vnhat).
\label{eq:aBlm}
\ea
An equivalent result is obtained for $\Del {}_{-1} \B$.  In general, the
multipole coefficients $a^{\B}_{lm}$ are complex, so that we can decompose
them into real and imaginary parts,
\be
a^{\B}_{lm} = a^{\B E}_{lm} + i\,a^{\B B}_{lm}.
\ee
One can easily show (\refapp{spinSH}) that under a change of parity
$a^{\B E}_{lm}$ transform as the spherical harmonic coefficients of a
vector (parity-odd), whereas $a^{\B B}_{lm}$, picking up an additional
minus sign, transforms as those of a pseudo-vector (parity-even).  
These thus correspond to the polar (``$E$'') and 
axial (``$B$'') parts of the vector $\B_i$.  

As required by parity, scalar perturbations do not contribute to
the axial part $a^{\B B}_{lm}$ (this is shown explicitly in \refapp{vector}).  
Thus, a measurement of the vector component $\B_i$ of standard ruler distortions
offers an additional possibility to probe tensor modes with large-scale
structure, as tensor modes do contribute to $a^{\B B}_{lm}$ (\refapp{vector}).  
Thus, in principle the axial component of $\B_i$ could
be of similar interest for constraining
tensor modes as weak lensing $B$-modes  \cite{paperII}, though one 
likely requires accurate redshifts to measure $\B_i$ to sufficient accuracy.  
We leave a detailed investigation of this for future work.  

The power spectrum of the $E$-mode of $\B$ due to standard
scalar perturbations is shown in \reffig{Cl} (see \refapp{calc}).  
While the dominant
contribution to $\C$ is $\propto k_\parallel^2/k^2\, \d^{\rm sc}_m(\vk,\zt)$
for a given Fourier mode of the matter density contrast in 
synchronous-comoving gauge (\refsec{C}), the corresponding contribution to $\B$ is 
$\propto k_\perp k_\parallel/k^2\, \d^{\rm sc}_m(\vk,\zt)$.  Even though
approximate scaling arguments suggest that $C_\C(l)$, $C^{EE}_{\B}(l)$
should scale roughly equally with $l$, we see that
$C_{\B}(l)$ scales faster with $l$ for $l \lesssim 500$.  The reason is 
that the projection kernel for the $E$-mode of $\B$ 
($\propto (\partial_x j_l)/x$) 
is relatively suppressed with respect to that of $\C$ ($\propto \partial_x^2 j_l$) at
large $x/l$.  Since $l \lesssim 500$ corresponds to a typical $k \lesssim 10^{-2} \iMpch$
at the source redshift, where $P_m(k) \propto k$, larger $x/l$ are favored
for progressively smaller $l$, leading to a more rapid decrease of $C_{\B}(l)$
towards smaller $l$.  This suppression is thus fundamentally a consequence of the shape
of the matter power spectrum.

\section{Transverse tensor: shear and magnification}

Finally, we have the purely transverse component,
\ba
\A_{ij} =\:& - \D\ln a \: \P_{ij}
- \frac12 \P_i^{\  k} \P_j^{\  l} h_{kl} \vs
& - \partial_{\perp\,(i} \D x_{\perp\,j)}
- \frac{1}{\chit} \D x_\parallel \P_{ij},
\label{eq:Aij}
\ea
where we have again inserted projection operators for clarity (note that
$\P_{ij}$ serves as the identity matrix on the sphere).  As
a symmetric matrix on the sphere, $\A_{ij}$ has a scalar component, given
by the trace $\A$, and two components of the traceless part which transform
as spin-2 fields on the sphere.  The trace corresponds to the change in
area on the sky subtended by two perpendicular standard rulers.  Thus,
it is equal to the magnification $\M$ (see also \reffig{SVT}).  
The two components of the traceless
part correspond to the shear $\g$.  If we choose a fixed coordinate
system $(\v{e}_\theta,\v{e}_\phi,\vnhat)$, we can thus write
\be
\A_{ij} = \left(\begin{array}{cc}
\M/2 + \g_1 & \g_2 \\
\g_2 & \M/2 - \g_1
\end{array}\right).
\label{eq:Aijcoord}
\ee
Below, we will derive magnification and shear without reference to a fixed
coordinate system.

\subsection{Magnification}
\label{sec:mag}

Taking the trace of \refeq{Aij} yields
\ba
\M \equiv\:& \P^{ij} \mathcal{A}_{ij} \vs
=\:& - 2\D\ln a 
- \frac12 \left(h^i_{\  i} - h_\parallel\right) 
+ 2\hat\k - \frac{2}{\chit} \D x_\parallel
\label{eq:mag}.
\ea
The magnification is directly related to the fractional perturbations in
distances (see \cite{HuiGreene,BonvinEtal06}) through 
\be
\frac{\Delta D_L}{D_L} = \frac{\Delta D_A}{D_A} = -\frac12 \M,
\ee
where the first equality for the luminosity distance follows from \refeq{DLDA}.  
The contributions to the magnification are straightforwardly interpreted
as coming from the conversion of coordinate distance to physical scale at
the source (from the perturbation to the scale factor $\D\ln a$ and the
metric at the source projected perpendicular to the line of sight, 
$h^i_{\  i}-h_\parallel$);  from the fact that the entire ruler is moved
closer or further away by $\D x_\parallel$;  and finally from the coordinate
convergence $\hat\k$ defined through
\be
\hat\k = -\frac12 \partial_{\perp\,i} \D x_\perp^i.
\ee
This term dominates the other contributions to $\M$ on small scales.  However,
the coordinate convergence is a gauge-dependent quantity; 
see for example App.~B2  in \cite{gaugePk}.  For the general
metric \refeq{metric} it is given by
\ba
\hat\k =\:& -\frac12\left[\frac12 \left((h^i_{\  i})_o - 3 (h_\parallel)_o \right)
- 2(B_\parallel - v_\parallel)_o
\right]  \label{eq:kappaG}\\
& +\frac12 \int_0^{\chit} d\chi\,\Bigg[
\partial^k_{\perp} B_k
- \frac{2}{\chi} B_\parallel
+ (\partial_\perp^l h_{lk}) \nhat^k \vs
& \hspace*{2cm} + \frac{1}{\chi} \left(h^i_{\  i} - 3 h_\parallel\right)
\vs
& \hspace*{2cm} + (\chit-\chi)\frac{\chi}{\chit}  \nabla_\perp^2\left\{
A - B_\parallel - \frac12 h_\parallel \right\}
\Bigg]. \nonumber
\ea
In conformal-Newtonian gauge, it
assumes its familiar form,
\ba
(\hat\k)_{\rm cN} =\:& - v_{\parallel o}
+ \frac12 \int_0^{\chit} d\chi\,\frac{\chi}{\chit} 
 (\chit-\chi)
\nabla_\perp^2 \left(\Psi - \Phi\right),
\label{eq:kcN}
\ea
with an additional term $-v_{\parallel o}$ contributing to the dipole
of $\hat\k$ only, which corresponds to the relativistic beaming effect
at linear order.  An explicit expression for the magnification 
in general gauge is straightforward to obtain, however it becomes lengthy.  
Here we just give the results for the synchronous-comoving and 
conformal-Newtonian gauges.  Using \refeq{DE} for synchronous-comoving gauge,
$(h^i_{\  i} - h_\parallel)/2 = 2 D - E_\parallel$, and we obtain  
\be
(\M)_{\rm sc} =- 2 (\D\ln a)_{\rm sc} - 2 D 
+ E_\parallel + 2 (\hat\k)_{\rm sc} - \frac2{\chit} \D x_\parallel.
\label{eq:magsc}
\ee
Since $(\D\ln a)_{\rm sc} = \d z$ defined in \cite{gaugePk}, we see
that we thus recover the covariant magnification, $\M = \d\M$, as derived
using an independent approach in \cite{gaugePk}.  

In conformal-Newtonian gauge [\refeq{metric_cN}], 
we have $(h^i_{\  i} - h_\parallel)/2 = 2\Phi$,
so that the magnification in this gauge becomes
\ba
\left(\M\right)_{\rm cN} =\:& \left[- 2 + \frac2{a H \chit}\right] (\D\ln a)_{\rm cN} - 2 \Phi  
+ 2(\hat\k)_{\rm cN} \vs
& - \frac{2}{\chit} \int_0^{\chit} d\chi\: (\Psi-\Phi)
+ \frac2{\chit} \int_0^{t_0} dt\: \Psi(\v{0},t)\,.
\label{eq:magcN}
\ea
The last term here is a pure monopole and thus usually absorbed in the ruler
calibration (since $r_0$ can rarely be predicted from first principles without
any dependence on the background cosmology).  Nevertheless, including this term ensures that gauge modes (for example superhorizon metric perturbations) do not affect the observed magnification.  In particular, in \refapp{magtest} we apply two
test cases to \refeq{magcN} where the monopole and dipole contributions (including $v_{\parallel o}$) become important.

\subsection{Shear}
\label{sec:shear}

We now consider the traceless part of $\A_{ij}$, given by
\ba
\g_{ij}(\vnhat) \equiv\:& \A_{ij} - \frac12\P_{ij} \M \vs
=\:& 
-\frac12 \left(\P_i^{\  k} \P_j^{\  l} 
- \frac12 \P_{ij} \P^{kl} \right) h_{kl} 
\vs
& - \partial_{\perp (i} \D x_{\perp\,j)} 
- \P_{ij} \hat\k.
\label{eq:shear1}
\ea
Here, the terms $\propto \P_{ij}$ in \refeq{Aij} drop out.  
The terms in the second line here is what commonly is regarded as the shear, 
i.e. the trace-free part of the transverse derivatives of the transverse
displacements.  The first term on the other hand is important to 
ensure a gauge-invariant result.  This is the term referred to as
``metric shear'' in \cite{DodelsonEtal}.  Its physical significance
becomes clear when constructing the Fermi normal coordinates for
the region containing the standard ruler.  

Consider a region of spatial extent $R$, say
centered on a given galaxy, with $R$
assumed to be much larger than the scale of individual galaxies.  
We can construct orthornomal Fermi normal coordinates \cite{Fermi,ManasseMisner}
around the center of this region, which follows a timelike geodesic,
by choosing the origin to be located at the center of the region at all times,
and the time coordinate to be the proper time of this geodesic.  
The spacetime in these Fermi coordinates $(t_F, x^i_F)$ then
becomes Minkowski, with corrections going as $x_F^2/R_c^2$ where $R_c$
is the curvature scale of the spacetime.  
Thus, as long as these corrections to the metric are negligible, 
there is no preferred direction in this frame, and the size of the standard ruler 
has to be (statistically) independent of the orientation.  The most 
obvious example is galaxy shapes, which are used for cosmic shear
measurements.  In the Fermi frame, galaxy orientations are random.  
Note that the Fermi coordinates are uniquely
determined up to three Euler angles.  The statement that galaxy orientations
are random in this frame is thus coordinate-invariant.  

As an example, consider the case where we have a purely spatial metric perturbation
(cf. \refeq{metric}) at a fixed time.  We can then expand around the
origin,
\be
h_{ij}(\vx) = h_{ij}(0) + h_{ij,k}(0) x^k.
\label{eq:gg}
\ee
Higher order terms are suppressed by $(x/R_c)^2$.  Now, consider coordinates
given by 
\be
a^{-1} x_{F}^i 
= x^i + \frac12 h_{ij}(0) x^j + 
\frac14 \left[2h_{ij,k}(0) - h_{jk,i}(0)\right] x^j x^k.
\label{eq:transf}
\ee
In these coordinates, the metric becomes 
\be
g^{F}_{\mu\nu} = \eta_{\mu\nu} + \O(x_{F}^2).  
\label{eq:gF}
\ee
Thus, it is in terms of the coordinates $x^i_{F}$ that galaxies
should be isotropically oriented on average, \emph{not} in terms
of the cosmological coordinates $x^i$.  Correspondingly, in order
to obtain the shear relative to the Fermi frame, we need to add the
transformation \refeq{transf} to the displacements $\D x^i$:
\be
\D x^i \rightarrow \D x^i + \frac12 h_{ij}(0) x^j 
+
\frac14 \left[2h_{ij,k}(0) - h_{jk,i}(0)\right] x^j x^k.
\ee
With these new displacements, the transverse derivative of the
transverse displacement becomes
\ba
\partial_{\perp (i} \D x_{\perp\,j)} \rightarrow \partial_{\perp (i} \D x_{\perp\,j)} + \frac12 \P_i^{\  k} \P_j^{\  k} h_{kl} + \O(h_{ij,k} x^k),
\label{eq:shearF}
\ea
where the last term is suppressed by the size of the ruler over the 
wavelength of the metric perturbation, and is thus negligible in the
small-ruler approximation.  We see that \refeq{shearF} agrees exactly
with the result derived above, \refeq{shear1} (after subtracting the
trace of \refeq{shearF}).  In other words,
the shear derived in the standard ruler formalism (\refsec{ruler}) is
equivalent to the statement that the ruler is isotropic in its Fermi
frame, the additional term coming from the transformation from global
coordinates to the local Fermi coordinates.  This additional term
was introduced in \cite{DodelsonEtal} as ``metric shear'', with a similar
motivation as given here.  In our case, this term is naturally included in the
standard ruler formalism.  

$\g_{ij}$ is a symmetric trace-free tensor on the sphere, and can thus
be decomposed into spin$\pm 2$ components (in analogy to the 
polarization of the CMB).  Following \refapp{spinSH} (see also \cite{Hu2000})
we can write $\g_{ij}$ as
\ba
\g_{ij} =\:& {}_2\g\, m_+^i m_+^j + {}_{-2}\g\, m_-^i m_-^j \vs
{}_{\pm 2}\g =\:&  m_\mp^i m_\mp^j \g_{ij},
\label{eq:sheardecomp}
\ea
where ${}_{\pm2}\g$ are spin$\pm2$ functions on the sphere (in analogy
to the combination of Stokes parameters $Q \pm i U$).  
We obtain for the shear components
\begin{widetext}
\ba
{}_{\pm 2}\g =\:& -\frac12 h_\pm - m_\mp^i m_\mp^j \partial_{\perp i} \D x_{\perp j} \vs
=\:& -\frac12 h_\pm  - \frac12  (h_{\pm})_o  - \int_0^{\chit} d\chi \Bigg[
\left(1 - 2\frac\chi{\chit}\right) 
\left[m_\mp^k \partial_\pm B_k
+ (\partial_\pm h_{lk}) m_\mp^l \nhat^k\right] 
- \frac{1}{\chit}h_{\pm}
\label{eq:shear2}\\
&\hspace*{4.2cm} +(\chit-\chi)\frac{\chi}{\chit} \Bigg\{
- m_\mp^i m_\mp^j \partial_i \partial_j A
+\nhat^k m_\mp^i m_\mp^j \partial_i \partial_j B_k
+ \frac12 m_\mp^i m_\mp^j(\partial_i \partial_j h_{kl})\nhat^k\nhat^l 
\Bigg\}\Bigg].
\nonumber
\ea
\refeq{shear2} is valid in any gauge.  We can now specialize to
the synchronous-comoving (sc) and conformal-Newtonian (cN) gauges:
\ba
\left({}_{\pm 2}\g\right)_{\rm sc} =\:& 
-\frac12 h_\pm - \frac12  (h_{\pm})_o - \int_0^{\chit} d\chi\, \Bigg[
\left(1-2\frac{\chi}{\chit}\right) (\partial_\pm h_{kl}) m_\mp^k \nhat^l
- \frac{1}{\chit} h_\pm
\label{eq:shear_sc}\\
& \hspace*{4.4cm} + (\chit-\chi)\frac{\chi}{\chit}
\frac12 (m_\mp^i m_\mp^j \partial_i\partial_j h_{lk})\nhat^l\nhat^k
\Bigg] \vs
\left({}_{\pm 2} \g\right)_{\rm cN} =\:& 
\int_0^{\chit} d\chi\, (\chit-\chi)\frac{\chi}{\chit}
 m_\mp^i m_\mp^j \partial_i\partial_j 
\left(\Psi - \Phi \right).
\label{eq:shear_cN}
\ea
\end{widetext}
In case of the cN gauge, we have used that $h_{ij} = 2\Phi \d_{ij}$,
and thus $h_\pm = 0$.  We see that \refeq{shear_cN} recovers the
``standard'' result; in other words, there are no additional
relativistic corrections to the shear \emph{in this gauge}.  This is not
surprising following our arguments above: in conformal-Newtonian gauge,
the transformation \refeq{transf} from global coordinates to the local Fermi
frame is isotropic since $h_{ij} = 2\Phi \d_{ij}$.  Thus, it does not
contribute to the shear.  Note however
that only scalar perturbations are included in this gauge;  when
considering vector or tensor perturbations, one has to use a different
gauge, for example synchronous-comoving gauge 
(see \cite{paperII} for a study of tensor perturbations).  Thus,
\refeq{shear2} and \refeq{shear_sc} are important new results.  

In \refapp{test}, we apply several test cases to the shear in 
synchronous-comoving gauge, \refeq{shear_sc}, in order
to verify that it is gauge-invariant and correctly reproduces known results.  
In particular, we consider a Bianchi~I cosmology which induces a shear
due to the anisotropic angular diameter distance.  We also show that
\refeq{shear_sc}, when restricted to scalar perturbations, does not
produce $B$-mode shear.  

\reffig{Cl} shows the angular power spectrum of shear and magnification
due to scalar perturbations
for a sharp source redshift $\zt = 2$ (see \refapp{calc}).  
For $l\gtrsim 10$, the results
follow the familiar relation $C_{\M}(l) = 4 C^{EE}_\g(l)$, valid when
all relativistic corrections to the magnification become irrelevant
so that $\M \simeq 2\hat\k$.  
These corrections slightly increase the magnification for small $l$.  
We also see that $\g$ and $\M$ are suppressed with respect to $\C$
and $\B$ (on smaller scales), at least when the latter are evaluated
for a sharp source redshift.  This is a well-known consequence of
the projection with the broad lensing kernel, leading to a cancelation
of modes that are not purely transverse (see e.g. \cite{JeongSchmidtSefusatti}).  

\section{Discussion}
\label{sec:disc}

Over the past decade, cosmology has benefited from a vast increase
in the available data, which have been exploited through a broad variety of methods
to constrain the history of structure in the Universe.  Clearly,
this calls for a rigorous investigation of what quantities precisely
are observable in the relativistic setting.  Some observables have
been investigated previously, most notably the number density of tracers
and the magnification.  Here, we have presented a unified relativistic analysis
of ``standard rulers'', where a standard ruler simply means there is
an underlying physical scale which we compare the observations to.  
This treatment applies to lensing measurements through galaxy ellipticities,
sizes and fluxes, or through standard candles, to distortions of 
cosmologial correlation functions,
and to lensing of diffuse backgrounds.   

We show that in this framework, for ideal measurements, one can
measure six degrees of freedom: a scalar corresponding to purely
line-of-sight effects; a vector (on the sphere) which corresponds
to mixed tranverse/line-of-sight effects; and a symmetric transverse
tensor on the sphere which comprises the shear and magnification.  
We obtain general, gauge-invariant expressions for the six observable
degrees of freedom, valid on the full sky.  These constitute the main
result of the paper and are given in \refeq{C}, (\ref{eq:Bpm}),
(\ref{eq:mag}), and (\ref{eq:shear2}).  
The vector component and the shear admit a decomposition into
$E/B$-modes.  The $B$-modes are free of all scalar contributions
(including lensing as well as redshift-space distortions) at the
linear level, making them ideal probes to look for tensor perturbations.  
As an application of our results, we study the shear induced by
tensor modes (gravitational waves) in \cite{paperII}.  

The logical next step is to construct estimators for these degrees of
freedom, based on measurements of the density field of tracers
(such as galaxies, the Lyman-$\alpha$ forest, 21cm emission, and so on).  
We will leave this for future work.

\acknowledgments
We would like to thank Yanbei Chen, Scott Dodelson, Olivier Dor\'e, 
Sam Gralla,
Chris Hirata, Wayne Hu, Bhuvnesh Jain, Marc Kamionkowski, Eiichiro Komatsu,
David Nichols, Samaya Nissanke, Enrico Pajer, and Matias Zaldarriaga for helpful discussions.  
FS thanks Masahiro Takada and the Kavli-IPMU, University of Tokyo,
for hospitality.  
FS is supported by the Gordon and Betty Moore Foundation at Caltech.

\begin{widetext}
\appendix

\section{Spherical harmonic decomposition of spin-$s$ functions}
\label{app:spinSH}

Here we outline our notation and useful results on the spherical harmonic
decomposition of tensors on the sphere.  
Throughout, latin indices $i, j,...$ denote components with respect
to Euclidean coordinates, and are raised and lowered with $\d_{ij}$.  
We follow standard convention, see \cite{Hu2000}. 
In particular, we do not include the Condon-Shortley phase 
in the spherical harmonics, so that $(Y_{lm})^* = Y_{l-m}$.  
Explicitly, in our convention the spherical harmonics are given by
\be
Y_{lm}(\theta,\phi) = \epsilon_m
\sqrt{\frac{2l+1}{4\pi}\frac{(l-|m|)!}{(l+|m|)!}}
P_l^{|m|}(\cos\theta) e^{im\phi},
\label{eq:Ylm}
\ee
where $\epsilon_m$ is a phase factor defined as
\be
\epsilon_m = \left\{\begin{array}{cl}
1&,~m>0 \\
(-1)^m     &,~m\le 0.
\end{array}\right.
\ee
We can define spin$\pm1$ unit basis vectors on the unit sphere
\ba
m^i_\pm \equiv\:& \frac{1}{\sqrt{2}} \left(e^i_\theta \mp i\,e^i_\phi\right)
= \frac{1}{\sqrt{2}}\left(\begin{array}{c}
\cos\theta \cos\phi \pm i \sin\phi\\
\cos\theta \sin\phi \mp i \cos\phi\\
-\sin\theta
\end{array}\right).
\ea
where $e_\theta,\,e_\phi$ are assumed orthonormal.  $m_{\pm}$ transform as 
spin$\mp1$ fields (see \refsec{intro}).  We have
\ba 
m^i_{\pm} m_{\pm i} =\:& 0, \qquad m^i_\pm m_{\mp i} = 1 \vs
m_\pm^i \nhat_i =\:& 0 \vs
\P^{ij} m_{\pm\,j} =\:& m_\pm^i.
\ea
Next, we define operators that raise and lower spin the spin $s$ of a
function (or tensor component) ${}_s f(\theta,\phi)$ through
\cite{GoldbergEtal,ZalSel97}
\ba
\Del\:{}_s f =\:& - \sin^s \theta \left[\frac{\partial}{\partial\theta}
+ \frac{i}{\sin\theta} \frac{\partial}{\partial\phi}\right] \sin^{-s}\theta
\:{}_s f \vs
\bar\Del\:{}_s f =\:& - \sin^{-s} \theta \left[\frac{\partial}{\partial\theta}
- \frac{i}{\sin\theta} \frac{\partial}{\partial\phi}\right] \sin^{s}\theta
\:{}_s f.
\label{eq:Deldef}
\ea
A straightforward calculation using partial integration shows that
\ba
\int (\Del {}_s f) {}_{s+1} g\: d\Omega =\:& \int (-\sin^s \theta)
\left(\left[\partial_\theta + i\sin^{-1}\theta \partial_\phi\right]
\sin^{-s}\theta\:{}_s f\right) {}_{s+1} g\:\sin\theta\, d\theta\, d\phi \vs
=\:& \int {}_s f \sin^{-s-1}\theta \left(\left[\partial_\theta + i \sin^{-1}\theta \partial_\phi \right] \sin^{s+1}\theta\:{}_{s+1} g\right)
\sin\theta\, d\theta\, d\phi \vs
=\:& \int {}_s f \left(-\bar\Del^* {}_{s+1} g\right) d\Omega,
\ea
in other words $-\bar\Del^*$ is the adjoint operator of $\Del$ with
respect to the standard measure on the sphere.   In many cases, we
will encounter functions given by ${}_s f(\theta,\phi) = e^{im\phi} {}_s\tilde f(\mu)$, where $\mu = \cos\theta$.  
In this case, \refeq{Deldef} simplifies to
\ba
\Del\:{}_s f =\:& -(1-\mu^2)^{(1+s)/2} \left[-\frac{\partial}{\partial\mu}
- \frac{m}{1-\mu^2}\right] (1-\mu^2)^{-s/2}
\:{}_s f \vs
\bar\Del\:{}_s f =\:& - (1-\mu^2)^{(1-s)/2} \left[-\frac{\partial}{\partial\mu}
+ \frac{m}{1-\mu^2} \right] (1-\mu^2)^{s/2} \:{}_s f,
\label{eq:Delmu}
\ea
which applied twice straightforwardly yields
\ba
\bar\Del^2\: {}_2 f(\mu,\phi) =\:& \left(-\frac{\partial}{\partial\mu} + \frac{m}{1-\mu^2}\right)^2 \left[ (1-\mu^2)\:{}_2 f(\mu,\phi) \right]
\vs
\Del^2\: {}_{-2} f(\mu,\phi) =\:& \left(-\frac{\partial}{\partial\mu} - \frac{m}{1-\mu^2}\right)^2 \left[ (1-\mu^2)\:{}_{-2} f(\mu,\phi) \right].
\label{eq:Del}
\ea

We can use the spin-raising/lowering operators to define spin-weighted spherical
harmonics through
\ba
{}_s Y_{lm} = \sqrt{\frac{(l-|s|)!}{(l+|s|)!}} 
\left\{
\begin{array}{cc}
\Del^s\:Y_{lm}, & s \geq 0 \vspace*{0.2cm}\\
(-1)^s \bar\Del^{|s|}\:Y_{lm}, & s < 0.
\end{array}
\right.
\label{eq:2Ydef}
\ea
Note again that our spherical harmonics are defined such that 
$\left[Y_{lm}\right]^* = Y_{l-m}$.  
\refeq{Deldef} together with \refeq{2Ydef} then
yield $[{}_sY_{lm}]^* = (-1)^{s}{}_{-s} Y_{l-m}$.  
For $s=\pm 2$, this is equivalent to the definition used
in \cite{Hu2000,Weinberg}:
\ba
{}_{\pm 2} Y_{lm} = 2\sqrt{\frac{(l-2)!}{(l+2)!}} m_\mp^i m_\mp^j \nabla_i\nabla_j Y_{lm}.
\ea
The spin-weighted spherical harmonics defined in \refeq{2Ydef} form
an orthornomal basis for spin-$s$ functions.  Using that $\Del^\dagger = -\bar\Del^*$, the orthonormality implies 
\be
\bar\Del^s \Del^s Y_{lm} 
= 
\Del^s \bar\Del^s Y_{lm} 
=
(-1)^s \frac{(l+|s|)!}{(l-|s|)!} Y_{lm}.
\ee
Returning to the decomposition
of a general spin$\pm s$ field (with $s > 0$), we can express the
components as
\ba
{}_{\pm s}A(\vnhat) =\:& \sum_{lm} a^A_{lm}\, {}_{\pm s}Y_{lm}(\vnhat),
\ea
Acting with the spin-lowering operator on ${}_{+s} A$ and vice versa then yields
\ba
\bar\Del^s {}_s A(\vnhat) =\:& \sum_{lm} a^A_{lm} \sqrt{\frac{(l-s)!}{(l+s)!}}
\bar\Del^s \Del^s Y_{lm}(\vnhat) = 
\sum_{lm} a^A_{lm} (-1)^s \sqrt{\frac{(l+s)!}{(l-s)!}} Y_{lm}(\vnhat)
\vs
\Del^s {}_{-s} A(\vnhat) =\:& \sum_{lm} a^A_{lm} 
(-1)^{s}
\sqrt{\frac{(l-s)!}{(l+s)!}}
\Del^s \bar\Del^s Y_{lm}(\vnhat) = 
\sum_{lm} a^A_{lm} 
\sqrt{\frac{(l+s)!}{(l-s)!}}
 Y_{lm}(\vnhat). 
\label{eq:spinalm}
\ea
We thus have
\ba
a^A_{lm} =\:& \int {}_{\pm s} A(\vnhat) \left[{}_{\pm s} Y_{lm}(\vnhat)\right]^* d\Omega\vs
=\:& \sqrt{\frac{(l-|s|)!}{(l+|s|)!}} (-1)^s \int \left[\bar\Del^s\: {}_{+s} A(\vnhat)\right] Y^*_{lm}(\vnhat) d\Omega
= \sqrt{\frac{(l-|s|)!}{(l+|s|)!}} 
\int \left[\Del^s\: {}_{-s} A(\vnhat)\right] Y^*_{lm}(\vnhat) d\Omega.
\label{eq:almspins}
\ea
This shows that the coefficients $a^A_{lm}$ have the desired
property of being invariant under a rotation
of the coordinate system around $\vnhat$.  We can then define
$E$- and $B$-components through
\ba
a^A_{lm} =\:& a^{AE}_{lm} + i a^{AB}_{lm} \vs
a^{AE}_{lm} =\:& \frac12\left(a^A_{lm} + a^{A*}_{lm} \right) \vs
a^{AB}_{lm} =\:& \frac1{2i}\left(a^A_{lm} - a^{A*}_{lm} \right).
\ea 
Under a parity transformation ($\vnhat \rightarrow \vnhat' = -\vnhat$),
$Y_{lm} \rightarrow (-1)^l Y_{lm}$.  
Hence, \refeq{2Ydef} yields 
${}_s Y_{lm} \to (-1)^{l} {}_{-s} Y_{lm} = (-1)^{l+s} [{}_s Y_{l-m}]^*$, 
and E- and B- components transform under parity as 
\ba
a^A_{lm} \rightarrow\:& (-1)^{l+s} a^{A*}_{l-m} \vs
a^{AE}_{lm} \rightarrow\:& (-1)^{l+s} a^{AE}_{l-m} \vs
a^{AB}_{lm} \rightarrow\:& - (-1)^{l+s} a^{AB}_{l-m}.
\label{eq:alm_parity}
\ea  
Thus, the $E$-component coefficients transform as expected of a spin-$s$
quantity derived from a scalar perturbation; for example, a vector given
by a gradient $\B_i = \partial_{\perp i} f$, whose components 
${}_{\pm1}\B$ transform as spin$\pm 1$ fields, is parity-odd
(just like the electric field).  On the other hand, the $B$-component
picks up an additional sign (parity-even, just like the magnetic field).  

\subsection{Angular power spectra of spin-$s$ functions}
\label{app:Cls}

The general procedure for obtaining the spherical harmonic coefficients
and angular power spectrum for a spin-$s$ quantity ${}_s A$ is as follows.  
The starting point is a relation between ${}_s A(\vnhat,\zt)$ and
the metric perturbations integrated over the (unperturbed) past light cone:
\be
{}_s A(\vnhat,\zt) = \int_0^{\chit} d\chi\: G[\chi, h(\chi\vnhat,\eta)],
\label{eq:step1}
\ee
where $\eta = \eta_0-\chi$, $h(\vx,\eta)$ stands for a single polarization 
state of any given metric perturbation, and
the kernel $G$ is a function of $\chi$ and $h$ and its derivatives.  
Throughout, we suppress the polarization index, although all polarization
states of course need to be summed in \refeq{step1}.  The goal is
to derive the contribution of a single Fourier mode of the metric perturbation,
and to subsequently add up the contributions of all Fourier modes.  For a 
scalar quantity ($s=0$), such as a density or temperature, this calculation is
straightforward since a scalar is invariant under a general rotation of the
coordinate system.  Thus, we can always align a given Fourier mode with the 
$z$-axis before summing up the contributions.  For a general spin-$s$ quantity
${}_s A$, this is not possible.  However, we can use the spin raising and 
lowering operators defined above to create a scalar quantity 
$\bar\Del^s\,{}_s A$ (for $s>0$) which allows us to easily sum up the contributions
of different Fourier modes.  The results of the previous section then
immediately tell us how the resulting spherical harmonic coefficients of
$\bar\Del^s\,{}_s A$ are related to those of ${}_s A$.

In detail, the calculation proceeds as follows:
\begin{enumerate}
\item Evaluate the contribution ${}_s A(\vnhat,\vk)$ from a single plane-wave
perturbation with a single circular polarization, with wavevector $\vk$ aligned with the
$z$-axis.   ${}_s A(\vnhat,\vk)$ is a function of $\vnhat$, usually
written in terms of $\mu \equiv \vnhat \cdot \hat{\v{z}}$ and azimuthal
angle $\phi$:
\be
{}_s A(\vnhat,\vk) = \int_0^{\chit} d\chi\:\tilde G(\chi, k, \mu) h(\vk) e^{i r\phi} e^{i x \mu},
\ee
where $h(\vk)$ is the Fourier amplitude of the mode at some reference 
epoch, $x = k\chi$ and $r$ is an integer.  In particular, $r=0$ if $h$ is a
scalar metric perturbation, $r=\pm 1$ for a vector perturbation, and $r=\pm2$ for
a tensor perturbation, depending on the polarization state.  $\tilde G$ is
an ordinary function obtained from $G[\chi, h]$ by replacing $\partial_i$
with $i k_i$, and pulling out $h(\vk)$.  Note that $\tilde G$ thus
contains the transfer function of the metric perturbation, relating
$h$ at the reference epoch to $h$ at conformal time $\eta = \eta_0-\chi$.  
\item Apply spin-raising or lowering operators to obtain a scalar quantity,
$\bar\Del^s\,{}_s A(\vnhat,\vk)$ if $s > 0$, $\Del^{|s|}\,{}_s A(\vnhat,\vk)$
if $s < 0$.  This quantity is a scalar on the sphere.  By virtue of the 
exponential $e^{i x \mu}$, we can turn derivatives with respect to $\mu$
into powers of $i x$, and powers of $\mu$ into derivatives with respect
to $ix$.  We can then write
\be
\bar\Del^s\,{}_s A(\vnhat,\vk) = \int_0^{\chit} d\chi\:\sum_i W_i(\chi) \hat Q_i(x) (1-\mu^2)^{|r|/2} e^{i r\phi} e^{i x \mu}\, h(\vk),
\label{eq:step2}
\ee
where $\hat Q_i(x)$ are derivative operators in $x$, $W_i(\chi)$ are
coefficient functions, and we have pulled out a factor of
$(1-\mu^2)^{|r|/2}$ for later convenience.  Note that since the $\hat Q_i(x)$
are constructed out of powers of $i k = i x/\chi$ and $\partial/\partial(i x)$, the
terms involving even powers of $x,\,\partial_x$ are real, while those
involving odd powers are imaginary.  Hence, $\hat Q_i^*(x) = \hat Q_i(-x)$.  
\item Since the angular dependence is now entirely in the factor $(1-\mu^2)^{|r|/2} e^{i r \phi} e^{i x \mu}$, we can straightforwardly expand this scalar quantity in terms 
of the standard spherical harmonics $Y_{lm}$ following \refeq{almspins}.  
The following relation which we prove in \refapp{proof} is useful:
\ba
\int d\Omega\: Y^*_{lm} (1-\mu^2)^{|r|/2} e^{ i r \phi} e^{i x \mu} 
=\:& \sqrt{4\pi (2l+1)} \sqrt{\frac{(l+|r|)!}{(l-|r|)!}}\, i^r\, i^l \frac{j_l(x)}{x^{|r|}} \d_{m r},
\label{eq:Ybessel}
\ea
where $r$ is an integer.  With this, we obtain
\ba
a^A_{lm}(k) = \d_{m r} \sqrt{4\pi (2l+1)} \sqrt{\frac{(l-|s|)!}{(l+|s|)!} \frac{(l+|r|)!}{(l-|r|)!}}
(-1)^s\, i^l 
\int_0^{\chit} d\chi \sum_i W_i(\chi) i^r\,\hat Q_i(x) \frac{j_l(x)}{x^{|r|}}
\,h(\vk) .
\label{eq:almk}
\ea
\item Following \refeq{alm_parity}, we can now separate the
$E$- and $B$-mode contributions:
\ba
a^{AE}_{lm}(k) =\:& \d_{m r} \sqrt{4\pi (2l+1)} \sqrt{\frac{(l-|s|)!}{(l+|s|)!} \frac{(l+|r|)!}{(l-|r|)!}}
(-1)^s\, i^l 
\int_0^{\chit} d\chi \sum_i W_i(\chi) i^r\,\Re\hat Q_i(x) \frac{j_l(x)}{x^{|r|}} 
\,h(\vk)
\label{eq:almkE}\\
a^{AB}_{lm}(k) =\:& \d_{m r} \sqrt{4\pi (2l+1)} \sqrt{\frac{(l-|s|)!}{(l+|s|)!} \frac{(l+|r|)!}{(l-|r|)!}}
(-1)^s\, i^l 
\int_0^{\chit} d\chi \sum_i W_i(\chi) i^r\,\Im\hat Q_i(x) \frac{j_l(x)}{x^{|r|}}
\,h(\vk).
\label{eq:almkB}
\ea
These constitute the multipole coefficients of $E$- and $B$-modes of ${}_s A$.  
\item The angular power spectra are straightforwardly obtained by
taking the expectation value of quadratic combinations of $a^{AX}_{lm}(k)$,
where $X = E, B$,
summing over $m$, and integrating over $(2\pi)^{-6} d^3k\, d^3k'$.  
Note that since we derived the multipole coefficients through the scalar
quantity $\bar\Del^s\,{}_s A(\vnhat,\vk)$ which is invariant under
rotations of the coordinate system, we can always align the Fourier
mode with the $z$-axis, so that \refeqs{almk}{almkB} remain valid.
\ba
C_A^{XX}(l) =\:& \frac{1}{2l+1} \sum_{m=-l}^l\int \frac{d^3k}{(2\pi)^3}
\int \frac{d^3k'}{(2\pi)^3} \< a^{AX}_{lm}(k) a^{AX\,*}_{lm}(k')\> \vs
=\:& \frac2\pi \frac{(l-|s|)!}{(l+|s|)!} \frac{(l+|r|)!}{(l-|r|)!} N_P  \int k^2 dk P_h(k) |F_l^X(k)|^2 \label{eq:CAXX}\\
F_l^E(k) =\:& \int_0^{\chit} d\chi \sum_i W_i(\chi) \Re \hat Q(k\chi) 
\frac{j_l(k\chi)}{(k\chi)^{|r|}} \vs
F_l^B(k) =\:& \int_0^{\chit} d\chi \sum_i W_i(\chi) \Im \hat Q(k\chi) 
\frac{j_l(k\chi)}{(k\chi)^{|r|}}.
\ea
Here, $P_h(k)$ is the power spectrum of $h(\vk)$ at the chosen reference
epoch, $N_P$ denotes the number of polarization states, and we have assumed
that the different polarization states have independent phases and equal
power spectra.  
\end{enumerate}
\refeq{CAXX} is a general expression for the $E/B$-mode
angular power spectra of a spin$-s$ observable induced by a spin$-r$ 
metric perturbation, which is straightforward to evaluate once an expression
of the form \refeq{step2} is given.  
Note that for $r=s=0$ and $N_P=1$, we recover the usual result for
scalar observables induced by scalar perturbations.  
In the following, and in the related papers \cite{paperI,paperII}, we will 
apply this result for $s=0,\pm1$ and $\pm 2$, as well as $r=0, \pm 2$.  

\subsection{Proof of \refeq{Ybessel}}
\label{app:proof}

The useful relation \refeq{Ybessel} is easily proven by induction over $r$.  
First, the case $r=0$,
\ba
\int d\Omega\: Y^*_{lm} e^{i x \mu} 
=\:& \sqrt{4\pi (2l+1)}  i^l j_l(x)\d_{m 0},
\ea
follows immediately from our definition of the spherical
harmonics, \refeq{Ylm}, and the partial wave expansion
\be
e^{ix\cos\theta} = \sum_l (2l+1) i^l j_l(x) P_l(\cos\theta).
\ee
Further, the definition of spherical harmonics, \refeq{Ylm}, yields
\ba
\int d\Omega\: Y^*_{lm} (1-\mu^2)^{|r|/2} e^{ i r \phi} e^{i x \mu} 
=\:& \epsilon_r \sqrt{4\pi(2l+1)} \sqrt{\frac{(l-|r|)!}{(l+|r|)!}} \d_{rm}
I^{|r|}_l(x) \vs
I^{|r|}_l(x) =\:&  \frac12\int_{-1}^1 d\mu (1-\mu^2)^{|r|/2} P^{|r|}_l(\mu) e^{ix\mu},
\label{eq:Idef}
\ea
where 
\be
P^m_l(x) = (-1)^m (1-x^2)^{m/2} \frac{d^m}{dx^m} P_l(x) \qquad (m \geq 0)
\ee
is the associated Legendre polynomial.  
Comparing with \refeq{Ybessel}, the conjecture to prove is thus
\ba
I^{|r|}_l(x) 
= 
\frac{(l+|r|)!}{(l-|r|)!}
i^{|r|+l} \frac{j_l(x)}{x^{|r|}}
\label{eq:conj}
\ea
We now proceed the proof by induction, assuming that \refeq{conj} holds
for some $r > 0$ (without loss of generality).  
Using one partial integration on $I^{r+1}_l(x)$ we obtain
\ba
I^{r+1}_l(x) =\:& 
\frac12 \int_{-1}^1 d\mu (1-\mu^2)^{(r+1)/2} 
(-1)^{(r+1)} (1-\mu^2)^{(r+1)/2} \left[\frac{d^{r+1}}{d\mu^{r+1}} P_l(\mu)\right] e^{ix\mu}
\vs
=\:& 
\frac12\int_{-1}^1 d\mu (-1)^{r} \left[\frac{d^{r}}{d\mu^{r}} P_l(\mu)\right]
\frac{d}{d\mu}\left[ (1-\mu^2)^{r+1} e^{ix\mu} \right]
\vs
=\:& 
\frac12\int_{-1}^1 d\mu (-1)^{r} (1-\mu^2)^r \left[\frac{d^{r}}{d\mu^{r}} P_l(\mu)\right]
\left[ -2\mu (r+1) + (1-\mu^2) i x  \right] e^{ix\mu}
\vs
=\:& \left[ 2 (r+1) \partial_x + x (1 + \partial_x^2)  \right]  
i I^r_l(x)
\ea
where in the third line we have converted powers of $i\mu$ to
derivatives $\partial_x$.
Now we use that, by assumption, \refeq{conj} holds for $r$, which leads to
\ba
I_l^{r+1}(x)
=\:& \frac{(l+r)!}{(l-r)!}
i^{r+1+l} \left[ 2 (r+1) \partial_x + x (1 + \partial_x^2)  \right]  
\frac{j_l(x)}{x^r}.
\ea
Straightforward 
algebra, together with the differential equation satisfied by
spherical Bessel functions, $j_l'' = -2j_l'/x + [l(l+1)/x^2-1] j_l$ leads to
\be
I_l^{r+1}(x) 
= 2 \frac{(l+r+1)!}{(l-r-1)!} i^{r+l+1} \frac{j_l(x)}{x^{r+1}},
\ee
which proves the conjecture \refeq{conj}. \hfill$\square$

\section{Geodesic equation and displacements}
\label{app:geod}

In this appendix, we outline the derivation of the displacements
\refeqs{Dxpar}{Dlna} in the general gauge given by \refeq{metric} and
\refeq{umu}.  This is a generalization of the derivation in \cite{gaugePk}
(who considered synchronous-comoving gauge), and a special case
of the treatment in \cite{PyneBirkinshaw}.  
Choosing the (zeroth order) comoving distance as affine parameter,
the photon momentum can be written as
\be
\frac{dx^\mu}{d\chi} = (-1+\d\nu,\: \nhat^i + \d e^i).
\label{eq:phmom}
\ee
The observer coincides with the location of the photon at $\chi=0$ and is assumed to follow a geodesic comoving with the cosmic fluid.  We choose him or her to lie at the spatial origin $\v{0}$, and to observe at a \emph{proper time} $t_0$.  The distinction between a fixed (gauge-dependent) coordinate time and a fixed proper time of observation affects the monopole of some of the standard ruler observables.  The geodesic equation then becomes
\ba
\frac{d}{d\chi}\frac{dx^\mu}{d\chi} + \Gamma^\mu_{\alpha\beta} 
\frac{dx^\alpha}{d\chi}\frac{dx^\beta}{d\chi} =\:0.
\ea
The zero-th order parts just yield $dx^\mu/d\chi = $const. 
We now turn to the first order part.  
The temporal component gives
\ba
\frac{d}{d\chi} \d\nu + \Gamma^0_{\alpha\beta} 
\frac{dx^\alpha}{d\chi}\frac{dx^\beta}{d\chi} =\:0 \vs
\frac{d}{d\chi} \d\nu + A'
- 2 A_{,i} \nhat^i + \left(\frac12 h_{ij}' + B_{(i,j)}\right) \nhat^i \nhat^j =\:0 \vs
\frac{d}{d\chi} \left(\d\nu - 2 A\right) = A' 
- \frac12 h_{\parallel}' - \partial_\parallel B_\parallel.
\ea
The spatial components yield
\ba
\frac{d}{d\chi} \d e^i + \Gamma^i_{\alpha\beta} 
\frac{dx^\alpha}{d\chi}\frac{dx^\beta}{d\chi} =\:0 \vs
\frac{d}{d\chi} \d e^i + A^{,i} - B^i{}'
- 2 \frac12 h'^{i}_{\  j} \nhat^j -2 \frac12 (B_{j,i}-B_{i,j})\nhat^j
+\frac12\left(h^i_{\  j,k} + h^i_{\  k,j} - h_{jk}^{\  \  ,i}\right) \nhat^j\nhat^k =\:0 \vs
\frac{d}{d\chi} \left(\d e^i + B^i + h^i_{\  j}\nhat^j \right) = 
- A^{,i} + \partial_i B_\parallel - B_{\perp i} + \frac12 h_\parallel^{\  ,i} - \frac1\chi \P^{ij} h_{jk} \nhat^k.
\ea
Next, we need to obtain the initial conditions at the observer 
for the quantities $\d\nu$, $\d e^i$.  
For this, we consider an orthonormal tetrad $(e_a)^\mu$, defined through
\be
g_{\mu\nu} (e_a)^\mu (e_b)^\nu = \eta_{ab}.  
\ee
Here, we use $a,b=0,1,2,3$ for the space-time index of the tetrad
$(e_a)$, while $\mu,\nu=0,1,2,3$ denotes the coordinate index of the tetrad $(e_a)^\mu$.
To zeroth order, we set $(e_b)^\mu = a^{-1}\: \d_{b}^{\  \mu}$.  
Using the perturbed FRW metric \refeq{metric}, this yields at first order
\ba
-1 =\:& g_{\mu\nu} (e_0)^\mu (e_0)^\nu = - a^2 (1+2A) [(e_0)^0]^2 \vs
\d_{ij} =\:& g_{\mu\nu} (e_i)^\mu (e_j)^\nu = a^2 (e_i)^k (e_j)^l (\d_{kl} + h_{kl})
\vs
0 =\:& g_{\mu\nu} (e_0)^\mu (e_i)^\nu = a^2 \left[-(e_0)^0 (e_i)^ 0 - B_j (e_i)^j (e_0)^0 
+ (e_0)_j (e^i)^j\right] .
\ea
The first line yields $(e_0)^0 = a^{-1} (1 - A)$, while the second line yields
$(e_k)^l = a^{-1} \left(\d_k^{\  l} - h_k^{\  l}/2\right)$.  
With this, the third line becomes
\be
0 = - a (e_i)^0 - B_i + a (e_0)_i.
\label{eq:int1}
\ee
Further, we require that the spatial hypersurfaces spanned by $(e_i)^\mu$ 
be orthogonal to the four-velocity of comoving observers (\refeq{umu}):
\be
0 = (e_i)^\mu\,u_\mu = - (e_i)^0 a(1+A) + v_i - B_i
\ee
Hence, we have $(e_i)^0 = a^{-1} (v_i-B_i)$, and \refeq{int1} yields 
$(e_0)^i = a^{-1} v^i$.  We thus have
\ba
(e_0)^\mu =\:& a^{-1} \left(1-A,\, v^i \right) \vs
(e_j)^\mu =\:& a^{-1} \left(v_j-B_j,\, \d_j^{\  i} - \frac12 h_j^{\  i}\right).
\label{eq:tetrad}
\ea
Comparing the first line with \refeq{umu}, we indeed see that
$(e_0)^\mu = u^\mu$, as desired.  We can also straightforwardly derive 
$(e_a)_\mu = g_{\mu\nu} (e_a)^\nu$, leading to
\ba
(e_0)_\mu =\:& a \left(-1-A,\, v_i-B_i \right) \vs
(e_j)_\mu =\:& a \left(-v_j, \d_{ji} + \frac12 h_{ji} \right).
\ea
We now require that the normalized photon momentum at the observer, 
$\hat p^\mu = (-1+\d\nu_o, \nhat^i + \d e^i_o)$, measured with respect to this
tetrad has the components $(1,\,\nhat^i)$ (note that for our choice of
affine parameter, the photon momentum is past-directed).  We obtain
\ba
1 =\:& \left[a^{-2}\, (e_0)_\mu\: p^\mu\right]_o = \left[ a^{-2}\, u_\mu p^\mu\right]_o = 1 - \d a_o + A_o - \d\nu_o + (v_{i\, o} - B_{i\, o})\nhat^i
\vs
\nhat_i =\:& \left[a^{-2}\, (e_i)_\mu p^\mu\right]_o = \left(1 -\d a_o \right)\nhat_i + v_{i\,o} + \d e_{i\,o} + \frac12 (h_{ij})_o \nhat^j,
\label{eq:pobs}
\ea
where the $a^{-2}$ factors come from the transformation of the affine parameter with respect to the comoving metric, $\chi$, to that corresponding to the physical metric, $\lambda$, through $d\chi/d\lambda = a^{-2}$ \cite{gaugePk}.  Further, we have included the perturbation of the scale factor at observation $\d a_o = a_o -1$, since we assume that the proper time of the observer at observation $t_0$ fixes the choice of scale factor via the relation $a(t_0) = 1$ in the background.  \refeq{pobs} yields
\ba
\d\nu_0 =\:& -\d a_o + A_o + v_{\parallel o} - B_{\parallel o} \vs
\d e^i_o =\:& \d a_o \nhat^i - v^i_o - \frac12 (h^i_{\  j})_o \nhat^j.
\ea
Note the aberration term $-v^i_o$ in $\d e^i_o$.  
We can now integrate the geodesic equations given these initial
conditions:
\ba
\d\nu(\chi) =\:& 2 A(\chi) + \int_0^\chi d\chi'\left[
A' - \frac12 h_{\parallel}' - \partial_\parallel B_\parallel\right] + {\rm const.}\vs
=\:& -\d a_o - A_o + v_{\parallel o} - B_{\parallel o} + 2 A(\chi) + \int_0^\chi d\chi'\left[
A' - \frac12 h_{\parallel}' - \partial_\parallel B_\parallel\right].
\label{eq:dnu}
\ea
The spatial component yields
\ba
\d e^i(\chi) =\:& \d a_o \nhat^i
+ \frac12 (h^i_{\  j})_o\, \nhat^j -v^i_o + B^i_o - B^i(\chi) - h^i_{\  j}(\chi)\nhat^j
+ \int_0^\chi d\chi' \left[
- A^{,i} 
+\nhat^jB_{j,i}
+ \frac12 h_{jk}^{\  \  ,i} \nhat^j\nhat^k
\right].
\label{eq:deiG}
\ea
Integrating again yields the temporal and spatial displacements:
\ba
\d x^0(\chit) =\:& \int_0^{\chit} d\chi\: d\nu(\chi) + \rm const\vs
=\:& \left[-\d a_o - A_o + v_{\parallel o} - B_{\parallel o}\right]\chit 
+ \int_0^{\chit} d\chi\left[ 2 A(\chi) + (\chit-\chi) \left\{
 A' - \frac12 h_{\parallel}' - \partial_\parallel B_\parallel\right\}
\right] + \rm const \vs
\d x^i(\chit) =\:& \int_0^{\chit}d\chi\: \d e^i(\chi) + {\rm const} \vs
=\:& \left[\d a_o \nhat^i + \frac12 (h^i_{\  j})_o\, \nhat^j + B^i_o - v^i_o\right] \chit 
+ \int_0^{\chit} d\chi \left[
- B^i - h^i_{\  j}\nhat^j + (\chit-\chi)\left\{
- A^{,i} +\nhat^jB_{j,i}
+ \frac12 h_{jk}^{\  \  ,i} \nhat^j\nhat^k
\right\}\right] \vs
& + \rm const\,.
\label{eq:dxiG}
\ea
The proper time $t_F$ of the observer is given in terms of the coordinate $\eta_o$ by (see \cite{Tpaper} for details)
\ba
t_F|_{\eta_o,\v{0}} =\:& \int_0^{\eta_o} \left[1 + A(\v{0},\eta') \right] a(\eta') d\eta' 
\vs
=\:& \bar t(\eta_o) + \int_0^{\bar t(\eta_o)}  A(\v{0},\bar\eta(t))\,  dt \,,
\label{eq:tFt0}
\ea
where $\bar t(\eta) = \int_0^{\eta} a(\eta')d\eta'$ is the proper time -- conformal time relation in the background Universe.  Requiring the observer to be at a 
fixed proper time, i.e. $t_F|_{\eta_o,\v{0}} = t_0$, and solving for the 
conformal time at observation $\eta_o$ yields at leading order
\be
\eta_o = \bar\eta(t_0) - \frac1{a_o}\int_0^{t_0} A(\v{0},\bar\eta(t)) dt
= \bar\eta(t_0) - \int_0^{t_0} A(\v{0},\bar\eta(t)) dt\,,
\label{eq:DeltaEta0}
\ee
where we have used that $a_o = 1$ at this order. 
This relation provides the boundary condition at the observer's 
location for the 0-component of $\d x^\mu$ through
\be
\d x^0(\chi=0) = \eta_o - \bar\eta(t_0) = - \int_0^{t_0} A(\v{0}, \bar\eta(t)) dt
\ee
so that
\be
\d x^0(\chit) = \left[-\d a_o - A_o + v_{\parallel o} - B_{\parallel o}\right]\chit 
 - \int_0^{t_0} A(\v{0}, t) dt + \int_0^{\chit} d\chi\left[ 2 A(\chi) + (\chit-\chi) \left\{
 A' - \frac12 h_{\parallel}' - \partial_\parallel B_\parallel\right\}
\right] \;. \label{eq:dx0G}
\ee
Next, we need to evaluate the scale factor and affine parameter at emission,
by requiring that the observed photon frequency match the redshift $\zt$.  
Since the photon momentum is given by \refeq{phmom}, we have
\be
1 + \zt = \frac{(a^{-2} u_\mu dx^\mu/d\chi)_e}{(a^{-2} u_\mu dx^\mu/d\chi)_o}
= \frac{a^{-1}(x^0) (1 + A - \d\nu + v_\parallel - B_\parallel)_e}
{a_o^{-1}(1 + A - \d\nu + v_\parallel - B_\parallel)_o},
\label{eq:zmatch}
\ee
where a subsript $e$ denotes the emission point, 
and we have decomposed $B_i = B_\parallel \nhat_i + B_{\perp\,i}$.  
The initial conditions for $\d\nu$ imply that the denominator is 1, and
we obtain
\be
\frac{a(x^0)}{\tilde a} = 1 + A - \d\nu + v_\parallel - B_\parallel
\equiv 1 + \D\ln a,
\label{eq:ax0}
\ee
where all quantities on the right-hand side are evaluated at emission,
and we have defined the perturbation to the logarithm of the scale factor
as emission, $\D\ln a$.  Explicitly,
\ba
\D\ln a =\:&  A - \d\nu + v_\parallel - B_\parallel 
\vs
=\:& A_o - A + v_\parallel - v_{\parallel o} 
+ \int_0^{\chit} d\chi\left[
- A' + \frac12 h_{\parallel}' + B_\parallel'\right]
- H_0 \int_0^{t_0} A(\v{0},\bar\eta(t)) dt\,.
\label{eq:Dlna2}
\ea
Here, we have used that the perturbation in scale factor at observation from $a(t_0)=1$ is given through \refeq{tFt0} by
\be
\d a_o = - \frac{da}{dt}\Big|_o \int_0^{t_0} A(\v{0},\bar\eta(t)) dt
= - H_0 \int_0^{t_0} A(\v{0},\bar\eta(t)) dt\,.
\ee
Since $\tilde a = 1/(1+\zt)$, and $a(x^0) = 1/(1+\bar z)$ where $\bar z$
is the redshift one would observe for the same source in an unperturbed
Universe, we can write this as
\be
1+\zt = (1+\bar z) \left( 1 + \D\ln a\right).
\label{eq:zmatch2}
\ee
The perturbation to the conformal time at emission has two contributions,
from the temporal displacement $\d x^0$ and from the perturbation to
the affine parameter from its zeroth-order value, $\chit$:
\be
a(x^0) = \tilde a + \tilde a \frac{\partial\ln a}{\partial \eta} \left[
\d x^0 - \d\chi\right].
\ee
\refeq{ax0} thus yields
\be
\frac{a(x^0)}{\tilde a} = 1 + a H \left[\d x^0 - \d\chi \right] = 1 + \D \ln a\;,
\ee
so that
\be
\d\chi = \d x^0 - \frac{1+\zt}{H(\zt)}\D\ln a.
\ee
We can now assemble the total line-of-sight deflection:
\ba
\D x_\parallel =\:& \d x^i \nhat_i + \d \chi
= \d x_\parallel + \d x^0 - \frac{1+\zt}{H(\zt)}\D\ln a
\vs
=\:& -\int_0^{t_0} A(\v{0},t) dt + \int_0^{\chit} d\chi\left[ A - B_\parallel - \frac12 h_\parallel
\right]
- \frac{1+\zt}{H(\zt)} \D \ln a.
\label{eq:DxparG}
\ea
For the transverse deflection, we obtain
\ba
\D x_\perp^i =\:& \d x^i - \nhat^i \d x_\parallel \vs
=\:& \left[\frac12 \P^{ij} (h_{jk})_o\, \nhat^k + B^i_{\perp o} - v^i_{\perp o}\right] \chit 
+ \int_0^{\chit} d\chi \left[
- B_\perp^i - \P^{ij} h_{jk}\nhat^k + (\chit-\chi)\left\{
- \partial_\perp^i A + \nhat^k \partial_{\perp}^i B_k
+ \frac12 (\partial_\perp^i h_{jk})\nhat^j\nhat^k
\right\}\right],
\label{eq:DxperpG} 
\ea
which can be further manipulated to yield \refeq{Dxperp2}.

\section{Test cases for the magnification}
\label{app:magtest}

In this section we apply test cases to the magnification derived in \refsec{mag}.  
We have shown that for synchronous-comoving gauge, $\M$ is identical to the
magnification derived in \cite{gaugePk}, where a wide variety of test cases
has been applied to this result.  Thus, we primarily need to test the terms
involving perturbations to the temporal components of the metric.  We present
two test cases.  First, a spatially constant but time-dependent perturbation
to $g_{00}$.  Using a change of time coordinate, this metric can be transformed
into an FRW metric with perturbed scale factor.  Second, a pure-gradient metric
perturbation in an Einstein-de Sitter Universe in conformal-Newtonian gauge.  
We verify that this metric perturbation has no impact on the observed magnification.  
In both test cases, the terms ensuring a fixed proper time of observation will prove essential.  

\subsection{Perturbed expansion history}

Consider a perturbed FRW metric given by
\be
ds^2 = -[1 + 2\Psi(\tilde t)] d\tilde t^2 + \tilde a^2(\tilde t) d\tilde{\vx}^2\,.
\label{eq:FRWc}
\ee
Further, consider an observer in this Universe ignorant about the perturbation $\Psi(\tilde t)$.  
They measure an age of the Universe $t_0$, which corresponds to their
proper time since the Big Bang.  We assume that they define
spatial coordinates such that $\tilde a(t_0) = 1$.  Then, they
will assign an object with observed redshift $\zt$ a comoving 
distance given by
\be
\chit = \int_{\tilde t_e}^{t_0} \frac{d\tilde t}{\tilde a(\tilde t)};\quad
\tilde a(\tilde t_e) = (1+\zt)^{-1}\,.
\ee
We now define a new time coordinate
\be
t(\tilde t) = \tilde t + \int_0^{\tilde t} \Psi(\tilde t') d\tilde t' + d\,,
\ee
where $d$ is a constant.  Notice that the proper time $t_F$ for a comoving observer ($\vx=$~const) implied by \refeq{FRWc} is
\be
t_F(\tilde t) = \int_0^{\tilde t} [1+\Psi(\tilde t') d\tilde t'\,.
\ee
Hence, the new time coordinate $t$ is identical to the proper time up to an arbitrary constant.  In the following, we will set the constant $d$ to zero since it will not have any observable impact.  Thus, working to linear order in $\Psi$ as we do throughout, 
\be
\tilde t(t) = t - \int_0^t \Psi(t') dt'\,.
\ee
With this time coordinate, the metric becomes
\ba
ds^2 =\:& - dt^2 + a^2(t) d\tilde{\vx}^2 
\label{eq:FRWa}\\
a(t) \equiv\:& \tilde a[\tilde t(t)] = 
\tilde a(t) \left[1 - \tilde H(t) \int_0^t \Psi(t') dt'\right]\,. 
\label{eq:atransform}
\ea
Thus, \refeq{FRWc} describes a homogeneous FRW cosmology written with an unnatural time coordinate.  The magnification that the observer assigns to, say, a standard candle at observed redshift $\zt$ is then proportional to the fractional difference between $\chit$ and the \emph{actual} comoving distance in the unperturbed FRW universe, \refeq{FRWa}:
\be
\M = -2 \frac{\chi - \chit}{\chit}\,.
\ee
We thus need to calculate $\chi$ for this source, given the fixed proper time at observation $t_0$.  The coordinate time corresponding to this proper time is 
\be
\tilde t(t_0) = t_0 - \int_0^{t_0}\Psi dt'\,,
\ee
so that the \emph{actual} FRW scale factor at observation is
\be
a_o = \tilde a[\tilde t(t_0)] = \tilde a(t_0) \left[1 - \tilde H(t_0) \int_0^{t_0}\Psi dt' \right] = 1 - \tilde H_0 \int_0^{t_0}\Psi dt'\,, 
\ee
where we have defined $\tilde H_0 = H(t_0)$.  Since in the $t$-coordinate
system the metric describes an unperturbed FRW Universe, the scale factor
at emission is given by
\be
a_e = (1+\zt)^{-1} a_o = \tilde a_e \left[1 - \tilde H_0 \int_0^{t_0} \Psi dt'\right]\,.
\label{eq:ae_aetilde}
\ee
We can now evaluate the true comoving distance $\chi$.  Note that the observer defines their spatial coordinates $\tilde{\vx}$ with respect to some local rulers at the observation time.  We thus need $\chi$ to refer to the corresponding physical coordinates $\tilde{\vx}/a_o$, which yields
\ba
\chi = \int_{t_{\rm em}}^{t_0} \frac{dt}{a(t) / a_o} = \left[1 - \tilde H_0 \int_0^{t_0} \Psi dt'\right]
\int_{t_{\rm em}}^{t_0} \frac{dt}{a(t)}\,,
\ea
where $a(t_{\rm em}) = a_e$. Since $dt = (1+\Psi) d\tilde t$, we then have
\ba
\chi =\:& \left(1 - \tilde H_0 \int_0^{t_0} \Psi dt' \right) \int_{\tilde t(t_{\rm em})}^{\tilde t(t_0)} \frac{[1+\Psi(\tilde t)] d\tilde t}{a[t(\tilde t)]}\,.
\ea
We now need to derive $t_{\rm em}$, the time coordinate at emission in the $t$-coordinate system.  Using \refeq{atransform} and \refeq{ae_aetilde}, we express $a(t_{\rm {em}})$
in two different ways:
\be
a(t_{\rm em}) = \tilde a(t_{\rm em}) \left[1 - \tilde H_e \int_0^{t_{\rm em}} \Psi dt' \right] 
= a_e = \tilde a(\tilde t_e) \left[1 - \tilde H_0 \int_0^{t_0} \Psi dt'\right].
\ee
As $t_{\rm em}$ coincides with $\tilde t_e$ at zeroth order in $\Psi$, we write $t_{\rm em} = \tilde t_e + \D t_{\rm em}$, and find at linear order in $\D t_{\rm em}$ 
\ba
& \tilde a(\tilde t_e) \left[1 +\tilde H_e \D t_{\rm em} - \tilde H_e \int_0^{\tilde t_e} \Psi dt' \right] 
= \tilde a(\tilde t_e) \left[1 - \tilde H_0 \int_0^{t_0} \Psi dt'\right]
\ea
so that
\be
 \D t_{\rm em} = \int_0^{\tilde t_e} \Psi dt' - \frac{\tilde H_0}{\tilde H_e} \int_0^{t_0} \Psi dt'\,.
\ee
Since $a[t(\tilde t)] = \tilde a\{\tilde t[t(\tilde t)]\} = \tilde a(\tilde t)$, we obtain
\ba
\chi =\:& \left(1 - \tilde H_0 \int_0^{t_0} \Psi dt' \right) \int_{\tilde t_e}^{t_0} \frac{[1+\Psi(\tilde t)] d\tilde t}{\tilde a(\tilde t)}
+ \left[-1 + \frac{\tilde H_0}{\tilde a_e \tilde H_e} \right] \int_0^{t_0} \Psi dt' \vs
=\:& \chit + \int_0^{\chit} \Psi d\chi + \left[ - \tilde H_0\chit - 1 +   
\frac{\tilde H_0}{\tilde a_e \tilde H_e} \right] \int_0^{t_0} \Psi dt'\,.
\ea
In the second line we have used the fact that $\Psi$ only depends on time to write the integral $\int \Psi dt/a$ as an integral over $\chi$.  Thus,
\ba
-\frac12 \M =\:& \frac{\chi - \chit}{\chit} = \frac1{\chit}\int_0^{\chit} \Psi d\chi
+ \left[- H_0 -\frac1{\chit}  + \frac{\tilde H_0}{\tilde a_e \tilde H_e\chit} \right] \int_0^{t_0}\Psi dt'\,.
\label{eq:magc}
\ea
We now evaluate our expression for the magnification, \refeq{magcN}, for \refeq{FRWc}.  Using that
$\Phi$, $v$, and $(\hat\k)_{\rm cN}$ vanish, we obtain
\ba
\M =\:& \left[- 2 + \frac2{a H \chit}\right] (\D\ln a)_{\rm cN} 
- \frac{2}{\chit} \int_0^{\chit} d\chi\: \Psi(\eta)
+ \frac2{\chit} \int_0^{t_0} dt\: \Psi(\eta(t))\,,
\ea
where
\ba
(\D\ln a)_{\rm cN} =\:& \Psi_o - \Psi 
- \int_0^{\chit} d\chi \Psi'(\eta) - H_0 \int_0^{t_0} \Psi(\eta(t)) dt \vs
=\:& \Psi_o - \Psi 
+ \Psi\Big|^{\rm src}_o   - H_0 \int_0^{t_0} \Psi(\eta(t)) dt 
%
= - H_0 \int_0^{t_0} \Psi(\eta(t)) dt\,.
\ea
Thus,
\ba
-\frac12 \M =\:& \frac1{\chit}\int_0^{\chit} d\chi \Psi + \left\{\left[- 1 + \frac1{a H \chit}\right] H_0 - \frac1{\chit} \right\} \int_0^{t_0} dt\: \Psi(\eta(t))\,,
\ea
which matches our expected result \refeq{magc}.  

\subsection{Pure-gradient metric perturbation}

As second test case we consider a constant$+$pure-gradient metric perturbation in conformal-Newtonian gauge (since we are mainly interested in testing the perturbation to the time-time component of the metric).  Since constant and pure-gradient metric perturbations can be removed by a suitable coordinate transform, they should not leave any impact in an observable such as the magnification.  

While this result should hold in general, we will specialize to an Einstein-de Sitter (EdS) Universe where distance and growth calculations are particularly simple (for $\Lambda$CDM or a general FRW cosmology, this calculation can be done numerically when all monopole and dipole terms are kept).  In EdS, the linear growth factor becomes $D(a) = a$, and we have $\Phi = -\Psi$, $\Psi'=0$ and $H_0 t_0 = 2/3$.  In this case, the magnification becomes
\ba
\left(\M\right)_{\rm cN} \stackrel{\rm EdS}{=}\:& 
\left[- 2 + \frac2{a H \chit}\right] \left[
v_\parallel - v_{\parallel o} - \Psi + \Psi_o - \frac23 \Psi_o \right]
+ 2 \Psi  
+ 2 (\hat\k)_{\rm cN} - \frac{4}{\chit} \int_0^{\chit} d\chi \:\Psi
+ \frac23 \frac2{\chit H_0} \Psi_o \vs
\left(\D\ln a\right)_{\rm cN} \stackrel{\rm EdS}{=}\:& v_\parallel - v_{\parallel o}
- \Psi + \Psi_o - \frac23 \Psi(\v{0})\,.
\ea
We now consider a constant+pure gradient potential perturbation,
\be
\Psi(\vx,\eta) = \Psi_0 \left[1 + \vk\cdot\vx \right]\,,
\ee
where $\Psi_0$ and $\vk$ are constants.  As before, the observer is assumed to be at $\vx=0$ and to be comoving.  We then obtain
\ba
\v{v} =\:& -\frac23 a^{1/2} \frac{\vk}{H_0} \Psi;\quad
v_{\parallel o} = -\frac23 \frac{k_\parallel}{H_0} \Psi_0\,.
\ea
The EdS background yields
\ba
\chit =\:& \int_{\tilde a}^1 \frac{da}{a^2 H(a)} = \frac2{H_0} (1- \tilde a^{1/2}) ; \quad
\tilde a \tilde H = H_0 \tilde a^{-1/2} \,.
\ea
The convergence is given by
\ba
(\hat\k)_{\rm cN} =\:& \frac23 \frac{k_\parallel}{H_0} \Psi_0 + \Psi_0 \int_0^{\chit} d\chi \frac{\chi}{\chit}(\chit-\chi) \nabla_\perp^2 \left(\vk\cdot\vnhat \chi \right)\vs
=\:& \frac23 \frac{k_\parallel}{H_0} \Psi_0 -2 k_\parallel\Psi_0 \int_0^{\chit} d\chi \frac{\chi}{\chit}(\chit-\chi) \frac1\chi
= \left(\frac23 \frac{1}{H_0} - \chit\right) k_\parallel \Psi_0\,,
\ea
where we have used that for a pure dipole, $\nabla_\perp^2 (\vk\cdot\vx) = -2 \vk\cdot\vx/|\vx|^2$ whereas the monopole contribution vanishes.  The monopole $\O(k^0)$ contribution to $\M$ is then
\ba
\M \stackrel{k^0}{=}\:& 
\left[- 2 + \frac2{a H \chit}\right] \left[  - \frac23 \Psi_0 \right]
+ 2 \Psi_0  
- 4 \Psi_0
+ \frac23 \frac2{\chit H_0} \Psi_0 \vs
=\:& \Psi_0 \frac13\left[
4 - \frac4{ a H \chit} -6 + 4 \frac1{\chit H_0}
\right] \vs
%
%
=\:& \Psi_0 \frac13\left[
-2 + 4 \frac{1- a^{1/2}}{H_0 \chit} 
\right] = \Psi_0 \frac13\left[-2 + 4 \frac12 \right] = 0\,,
\ea
as desired.  Using that $v_\parallel - v_{\parallel o} = \left(1 - a^{1/2}\right) \frac23 \frac{k_\parallel}{H_0} \Psi_0$, the dipole component $\propto k_\parallel$ is obtained as
\ba
\M \stackrel{k^1}{=}\:& k_\parallel \Psi_0 \bigg\{
\left[- 2 + \frac2{a H \chit}\right] \left[(1-a^{1/2}) \frac2{3 H_0} - \chit \right]
+ 2 \chit - 2 \chit  + 2 \frac2{3H_0}
 - \frac{4}{\chit} \frac12 \chit^2 
 \bigg\} \vs
=\:& k_\parallel \chit \Psi_0 \bigg\{
\left[- 2 + \frac{2a^{1/2}}{ H_0 \chit}\right] \left[-\frac23\right]
 + \frac4{3H_0 \chit}
 - 2 \bigg\}
\vs
=\:& k_\parallel \chit \Psi_0 \frac13\bigg\{-2
+\frac4{H_0\chit} (1-a^{1/2}) \bigg\} = 0\,,
\ea
as desired.  As for the previous test, the constant observer terms in $\D x_\parallel$ and $\D\ln a$ obtained by enforcing a fixed proper time of observation are crucial for obtaining a vanishing monopole.  The dipole on the other hand tests the non-trivial velocity terms in $\D\ln a$ and $(\hat\k)_{\rm cN}$.

\section{Test cases for the shear}
\label{app:test}

In this appendix, we consider test cases in order to validate the
expression for the shear, \refeq{shear_sc} in \refsec{shear}.  
For a larger set of
test cases applied to scalar quantities such as the observed galaxy
density and the magnification, see Appendix C in \cite{gaugePk}.  

The first case is a metric perturbation corresponding to a pure gauge mode, i.e
\be
h_{ij}(\vx,\eta) = A_{ij} + B_{ijk} x^k,
\ee
where $A_{ij}$ and $B_{ijk}$ are constant and symmetric in $i$ and $j$.  Such
a metric perturbation can be obtained through a coordinate transform
$x^i \to \hat x^i = x^i - A_{ij} x^j/2 - B_{ijk} x^j x^k/4$.  
Choosing the observer to be at the origin, we have
\ba
h_{\pm\,o} =\:& A_{\pm} \vs
h_{\pm\,s} =\:& A_\pm + B_{\pm\,k} \tilde x^k \vs
\int_0^{\chit} \frac{d\chi}{\chit} h_\pm =\:& A_\pm + \frac12 B_{\pm\,k}\tilde x^k \vs
\ea
where $B_{\pm\,k} = m_\mp^i m_\mp^j B_{ijk}$.  
Since $\partial_i\partial_jh_{kl} = 0$
and $\partial_i h_{kl} = B_{kli} = \rm const$, \refeq{shear_sc} then
yields
\be
{}_{\pm 2} \g(\vnhat) = -\frac12 A_\pm - \frac12 A_\pm -\frac12 B_{\pm\,k}\tilde x^k + A_\pm + \frac12 B_{\pm\,k}\tilde x^k = 0.
\ee
Thus, a gauge mode does not contribute to the shear.  

Further possible test cases
are a perturbed expansion history and spatial curvature.  Both
cases correspond to isotropic Universes, and the observed shear 
should thus be zero.  The former case is described by a metric
perturbation of the form $h_{ij} = A(\eta) \d_{ij}$.  Since this
implies that $\partial_i h_{jk}=0$ and $h_{\pm} = 0$, \refeq{shear_sc}
implies ${}_{\pm 2} \g =0$.  Spatial curvature is decribed by
a metric perturbation $h_{ij} = -K/4\, x_k x^k \d_{ij}$.  In this case,
$h_\pm = 0$, $\partial_i h_{kl} = -K/2\, x_i \d_{kl}$, and 
$\partial_i\partial_j h_{kl} = -K/2\, \d_{ij} \d_{kl}$.  Thus,
\ba
(m_\mp^i m_\mp^j \partial_i \partial_j h_{kl}) \nhat^k\nhat^l =
-\frac{K}2 (m^i_\mp m_{\mp\,i})^2 = 0
\vs
\nhat^l m_\mp^k m_\mp^i \partial_i h_{kl} = -\frac{K}2 (m^i_\mp x_i)
\nhat_k m_\mp^k = 0.
\ea
This implies that ${}_{\pm 2} \g = 0$ for spatial curvature as well.  
There is however one test case where the shear is non-trivial, which we
will consider next.

\subsection{Bianchi I cosmology}
\label{app:BianchiI}

A Bianchi I cosmology is an anisotropically expanding Universe.  Following
\cite{gaugePk}, we choose the 3-axis to be unperturbed, while the 
scale factors along the 1- and 2-axes are perturbed in the following
way:
\ba
a_1(\eta) =\:& a(\eta) [1 + s_1(\eta) - s_3(\eta)] \vs
a_2(\eta) =\:& a(\eta) [1 + s_2(\eta) - s_3(\eta)] \vs
a_3(\eta) =\:& a(\eta),
\ea
where $s_1(\eta)+s_2(\eta)+s_3(\eta) = 0$, and $s_i(\eta_0)=0$.  Relaxing
either of these conditions leads to cases we have studied above
(perturbed expansion history and pure gauge mode).  
The non-zero components of $h_{ij}$ are then given by
\ba
h_{11} =\:& 2[s_1(\eta)-s_3(\eta)] \vs
h_{22} =\:& 2[s_2(\eta)-s_3(\eta)].
\ea
Let us consider two lines of sight close to the unperturbed 3-axis.  
Specifically, we consider photon 4-momenta at the observer given by
$p_\mu^{(1)} = (-1,-\varsigma,0,-1)$ and $p_\mu^{(2)} = (-1,0,-\varsigma,0,-1)$,
where $\varsigma$ is the infinitesimal angle with the 3-axis.  
By following back these geodesics, one can straightforwardly derive
the angular diameter distances along the 3-axis, for an object extended
along the 1- and 2-axes \cite{gaugePk}:
\ba
D_{A,{\rm phys},a}(\eta) =\:& 
a(\eta) (\eta_0-\eta)\Bigg[1 + s_a(\eta) - s_3(\eta) 
+ \frac2{\eta_0-\eta} \int_\eta^{\eta_0} d\eta' \left(s_3(\eta')-s_a(\eta')\right) \Bigg], 
\ea
where $a=1,2$, and $\eta$ is the conformal time of emission.  
The observed ellipticity of galaxies, designed as
estimator for shear, can be written as
\ba
\epsilon_1 =\:& \frac12 \frac{I_{11} - I_{22}}{I_{11}+I_{22}}
= \frac12 \frac{D_{A,{\rm phys},1}^{-2} - D_{A,{\rm phys},2}^{-2}}
{D_{A,{\rm phys},1}^{-2} + D_{A,{\rm phys},2}^{-2}} \vs
\epsilon_2 =\:& \frac{I_{12}}{I_{11}+I_{22}} = 0,
\ea
where $I_{ij}$ are the quadrupole moments of the galaxy's light distribution,
which scale as $D_{A,\rm phys}^{-2}$.  Other definitions are possible,
however all of them agree at linear order.  $\epsilon_2$ vanishes, since
a ray with $p_\mu = (-1,-\varsigma/\sqrt{2},\varsigma/\sqrt{2},-1)$, 
propagating at $+45^\circ$ azimuthal angle to the 1-axis, yields the
same angular diameter distance as a ray with $p_\mu = (-1,-\varsigma/\sqrt{2},-\varsigma/\sqrt{2},-1)$, propagating at $-45^\circ$ angle.  
Assuming that the galaxies are on average round ($\<\epsilon_i\>=0$),
i.e. that they are not directly influenced by the anisotropic expansion,
we obtain
\ba
\epsilon_1 =\:& \frac14 \left [-2 s_1 + 2 s_3 + 2 s_2 - 2s_3
- \frac4{\eta_0-\eta} \int_\eta^{\eta_0}d\eta' \left[-s_1(\eta')+s_2(\eta')\right]\right] \vs
=\:& \frac12\left[s_2(\eta) - s_1(\eta)\right] + \frac1{\eta_0-\eta}
\int_\eta^{\eta_0}d\eta' \left[s_1(\eta')-s_2(\eta')\right].
\label{eq:eps1}
\ea
The extra factor of $1/2$ is due to the sum of moments in the
denominator in the definition of $\epsilon_1$.  
In order to compare with \refeq{shear_sc}, we use
\ba
h_\pm =\:& \frac12 (h_{11} - h_{22} \pm 2i h_{12}) \vs
=\:& s_1(\eta) - s_2(\eta).
\ea
We thus have $h_{\pm\,o} = 0$, and \refeq{shear_sc} yields for the
shear along the 3-axis
\ba
\g_1\pm i \g_2 =\:& 
-\frac12 (s_1(\eta)-s_2(\eta))
+ \int_0^{\chit} \frac{d\chi}{\chit} \left[s_1(\eta(\chi)) - s_2(\eta(\chi))\right] \vs
=\:& \frac12 (s_2(\eta)-s_1(\eta))
+ \frac{1}{\eta_0-\eta}\int_\eta^{\eta_0} d\eta' \left[s_1(\eta') - s_2(\eta')\right],
\ea
where we have used that ${}_{\pm 2}\g = \g_1\pm i \g_2$ for the coordinates
chosen here (see \refeq{Aijcoord}).  Since this expression is real,
$\g_2=0$, and the result is equal to $\g_1$, which moreover agrees  
with the correct physical result for $\epsilon_1$, \refeq{eps1}.  

\subsection{Shear from scalar perturbations}
\label{app:scalar}

In the derivation leading to \refeq{shear_sc}, we have not made any
assumptions about metric perturbations except that 
$\d g_{00} = 0 = \d g_{0i}$ (synchronous-comoving gauge).  
As a cross-check, we now consider the case of scalar perturbations,
where we can write
\be
h_{ij} = 2 D \d_{ij} + 2\left(\partial_i\partial_j - \frac13 \d_{ij} \nabla^2\right) E
\ee
in terms of the 3-scalar perturbations $D$ and $E$ (see \cite{gaugePk}).  
We thus have $h_\pm = 2 m_\mp^i m_\mp^j E_{,ij}$ and obtain
\ba
{}_{\pm2} \g(\vnhat) =\:& 
- m_\mp^i m_\mp^j \left(E_{,ij}(o) + E_{,ij}(s)\right)
- \int_0^{\chit} d\chi \Bigg\{
(\chit-\chi)\frac{\chi}{\chit}
m_\mp^i m_\mp^j \left(D_{,ij} - \frac13 \nabla^2 E_{,ij} 
+ \partial_\parallel^2 E_{,ij}\right) \vs
& \hspace{6cm} + 2\left(1-2\frac{\chi}{\chit}\right) 
 m_\mp^i m_\mp^j \partial_\parallel E_{,ij}
- \frac2{\chit} m_\mp^i m_\mp^j E_{,ij}
\Bigg\}.
\label{eq:shear_scalar}
\ea
We now consider the contribution of a 
single plane wave along the $z$-axis,
\be
D(\vx,\eta) = D(\vk,\eta) e^{i\vk\cdot\vnhat\chi} = D(\vk,\eta) e^{i x\mu},
\ee
and similarly for $E$.  We can then replace
\be
m_\mp^i m_\mp^j \partial_i \partial_j \to -\frac12(1-\mu^2) k^2.
\ee
This yields
\ba
{}_{\pm2} \g(\vk,\vnhat) =\:& 
+ \frac12 (1-\mu^2) k^2 \left(E(\vk,\eta_0) e^{ix\mu}\Big|_{x=0}
+ E(\vk, \tilde\eta) e^{i\tilde x\mu} \right) \vs
& + \int_0^{\chit} d\chi \Bigg\{
(\chit-\chi)\frac{\chi}{\chit}\frac12 (1-\mu^2) k^2
\left(D(\vk,\eta) + \frac13 k^2 E(\vk,\eta) - \mu^2 k^2 E(\vk,\eta)
\right) \vs
& \hspace{1.5cm} + \left(1-2\frac{\chi}{\chit}\right) 
i (1-\mu^2)\mu k^3 E(\vk,\eta)
- k^2 \frac1{\chit} (1-\mu^2) E(\vk,\eta)
\Bigg\} e^{ix\mu},
\ea
where $\tilde\eta = \eta_0-\chit$, and $\tilde x = k\chit$.  
Note that a scalar perturbation produces equal amplitudes of ${}_{\pm 2} \g$: 
there is no preferred handedness for scalar modes.  
Correspondingly, since this expression is $\propto e^{im\phi}$ with $m=0$, 
the spin-lowering and spin-raising actions [\refeq{Del}] become equivalent.  
We have
\ba
\bar\Del^2 {}_2\g =\:& \Del^2 {}_{-2}\g = \frac{\partial^2}{\partial\mu^2} \left[
(1-\mu^2) {}_2\g(\vnhat,\vk)\right] \vs
=\:& 
\frac12 k^2 \left(E(\vk,\eta_0) \left[Q^S_1(x) x^2 e^{ix\mu}\right]_{x=0}
+ E(\vk, \eta) \left[Q^S_1(x) x^2 e^{ix\mu}\right]_{x=k\chit} \right) \vs
& + \int_0^{\chit} d\chi \Bigg\{
(\chit-\chi)\frac{\chi}{\chit}\frac12 k^2
\left( Q^S_1(x) D(\vk,\eta) + \frac13 Q^S_1(x) k^2 E(\vk,\eta) 
- Q^S_2(x) k^2 E(\vk,\eta)
\right) \vs
& \hspace{1.5cm} - \left(1-2\frac{\chi}{\chit}\right) \frac1\chi
Q^S_3(x) k^2 E(\vk,\eta)
- \frac1{\chit} Q^S_1(x) k^2 E(\vk,\eta)
\Bigg\} x^2 e^{ix\mu}\Bigg|_{x=k\chi}.
\ea
Using that
\ba
\frac{\partial^2}{\partial\mu^2}\left[(1-\mu^2)^2 e^{ix\mu}\right]
=& \frac{\partial^2}{\partial\mu^2}\left[(1+\partial_x^2)^2 e^{ix\mu}\right]
= - (1+\partial_x^2)^2 \left[ x^2 e^{ix\mu}\right],
\vs
\frac{\partial^2}{\partial\mu^2}\left[(1-\mu^2)^2 \mu e^{ix\mu}\right]
=&
= i\partial_{x}(1+\partial_x^2)^2 \left[ x^2 e^{ix\mu}\right],
\vs
\frac{\partial^2}{\partial\mu^2}\left[(1-\mu^2)^2 \mu^2 e^{ix\mu}\right]
=&
= \partial_{x}^2 (1+\partial_x^2)^2 \left[ x^2 e^{ix\mu}\right],
\ea
we obtain the operators $Q^S_i(x)$ for the scalar case:
\ba
Q^S_1(x) =\:& - (1+\partial_x^2)^2
\vs
Q^S_2(x) =\:& (1+\partial_x^2)^2 \partial_x^2
\vs
Q^S_3(x) =\:& x (1+\partial_x^2) \partial_x.
\ea
Note that all these operators are real.  Hence, following the general
derivation in \refapp{Cls}, there are no parity-odd
terms in the scalar contributions to the shear, and thus no $B$-modes
as expected.

\section{Vector}
\label{app:vector}

In this section, we follow the procedure described in \refapp{Cls}
in order to derive the multipole moments of the vector component
${}_{\pm 1}\B$ induced by scalar and tensor perturbations.  For scalar
perturbations, we work in conformal-Newtonian gauge; there is no
gauge ambiguity for tensor modes.  Our goal is to show that scalar
perturbations do not contribute to $a_{lm}^{\B B}$, while tensor
perturbations do contribute.  

\subsection{Scalar modes}

We begin with \refeq{Bpm_cN}:
\ba
({}_{\pm 1}\B)_{\rm cN} =\:& -v_\pm + \frac{1+\zt}{H(\zt)}\Bigg(-\partial_\pm \Psi + \partial_\pm [v_\parallel  - v_{\parallel o}]
+ \int_0^{\chit} d\chi\frac{\chi}{\chit} \partial_\pm(\Phi'-\Psi')
\Bigg),
\ea
and specialize to the case of a single plane wave along the $z$-axis.  
Further, we write $v_i = V_{,i}$.  Noting that
\ba
\partial_{\perp i} v_\parallel =\:& \partial_{\perp i} (\nhat^j \partial_j V)
= \nhat^j \partial_{\perp i}\partial_j V + \frac1\chi \partial_{\perp i }V \vs
\partial_{\perp i} v_{\parallel o} =\:& \frac1{\chit} (\partial_{\perp i} V)_o
\ea
this yields:
\ba
({}_{\pm 1}\B)_{\rm cN}(\vnhat,\vk) =\:& - i k_\pm V(\vk,\tilde\eta) e^{i \tilde x\mu}
+ \frac{1+\zt}{H(\zt)}\Bigg(-i k_\pm \Psi(\vk,\tilde\eta) e^{i\tilde x \mu} 
+ \left(\frac{i k_\pm}{\chit}- k_\pm k_\parallel\right) V(\vk,\tilde\eta) e^{i\tilde x \mu}
- \frac{i k_\pm}{\chit} V(\vk,\eta_0) e^{ix\mu}|_{x=0} \vs
& \hspace*{4.5cm} + \int_0^{\chit} d\chi\frac{\chi}{\chit} i k_\pm [\Phi'(\vk,\eta)-\Psi'(\vk,\eta)] e^{i x \mu}
\Bigg) \vs
=\:& \frac{\sqrt{1-\mu^2}}{\sqrt2} \Bigg[ i k V(\vk,\tilde\eta) e^{i \tilde x\mu}
+ \frac{1+\zt}{H(\zt)}\Bigg(i k \Psi(\vk,\tilde\eta) e^{i\tilde x \mu} 
+ \left(-i \frac{k}{\chit} + \mu k^2\right) V(\vk,\tilde\eta) e^{i\tilde x \mu}
+ \frac{i k}{\chit} V(\vk,\eta_0) e^{ix\mu}|_{x=0} \vs
& \hspace*{5.5cm}
- \int_0^{\chit} \frac{d\chi}{\chit} i x [\Phi'(\vk,\eta)-\Psi'(\vk,\eta)] e^{i x \mu}
\Bigg)\Bigg],\nonumber
\ea
where as before $x = k\chi$, $\tilde x = k\chit$, and 
$\tilde\eta = \eta_0-\chit$.  Further, we have used that $k_\parallel = \nhat^i k_i = \mu k$, and $k_\pm = m_\mp^i k_i = -\sqrt{1-\mu^2}\, k /\sqrt{2}$.  
We now apply the spin-lowering operator \refeq{Delmu} to ${}_1\B$.  
Note that since ${}_{\pm 1}\B(\vnhat,\vk)$ is $\phi$-independent, this is
identical to the expression for $\Del\,{}_{-1}\B$.
\ba
\bar\Del {}_1 \B(\vnhat,\vk) =\:& \partial_\mu \left[\sqrt{1-\mu^2}\, {}_1\B(\vnhat,\vk)\right] \vs
=\:& \frac{\partial_\mu}{\sqrt2} \Bigg\{ (1-\mu^2) \Bigg[ i k V(\vk,\tilde\eta) e^{i \tilde x\mu}
+ \frac{1+\zt}{H(\zt)}\Bigg(i k \Psi(\vk,\tilde\eta) e^{i\tilde x \mu} 
+ \left(\mu -i \frac{1}{\tilde x}\right)k^2  V(\vk,\tilde\eta) e^{i\tilde x \mu}
+ i \frac{1}{\tilde x}\ k^2  V(\vk,\eta_0) e^{i x \mu}|_{x=0} \vs
& \hspace*{6cm}
- \int_0^{\chit} \frac{d\chi}{\chit} i x [\Phi'(\vk,\eta)-\Psi'(\vk,\eta)] e^{i x \mu}
\Bigg)\Bigg]\Bigg\} \vs
=\:& 
\frac{V(\vk,\tilde\eta)}{\sqrt2 \chit}
\hat Q_{BS1}(\tilde x) e^{i \tilde x\mu}
\vs
& + \frac{1+\zt}{\sqrt2 H(\zt)}\Bigg\{
\frac1{\chit} \Psi(\vk,\tilde\eta) \hat Q_{BS1}(\tilde x) 
 e^{i\tilde x \mu}
+ k^2 V(\vk,\tilde\eta) \left[\hat Q_{BS2}(\tilde x) - \frac1{\tilde x^2}\hat Q_{BS1}(\tilde x) \right] e^{i\tilde x \mu}
\vs
& \hspace*{1.9cm}
 + \frac1{\chit^2} \left[\hat Q_{BS1}(x) e^{i x\mu}\right]_
{x=0}\!\! V(\vk,\eta_0)
- \int_0^{\chit} \frac{d\chi}{\chit} \hat Q_{BS1}(x) [\Phi'(\vk,\eta)-\Psi'(\vk,\eta)]e^{i x \mu}
\Bigg\},
\ea
where we have defined
\ba
\hat Q_{BS1}(x) =\:& -x^2 -2x \partial_x - x^2 \partial_x^2
\vs
\hat Q_{BS2}(x) =\:& 1 + x\partial_x + 3 \partial_x^2 + x\partial_x^3.
\ea
This expression is in the desired form, \refeq{step2}, and we see that
all coefficients are real.  Thus, scalar perturbations only contribute
to the (polar) $E$-component of $\B_i$.  

\subsection{Tensor modes}

We begin with \refeq{Bpm_sc}:
\ba
({}_{\pm1}\B)_{\rm sc} =\:& \frac{1+\zt}{2 H(\zt)}\int_0^{\chit}d\chi\frac{\chi}{\chit}\partial_\pm h_\parallel'
= \frac{1+\zt}{2 H(\zt)}\int_0^{\chit}d\chi\frac{\chi}{\chit}
\left( (\partial_\pm h'_{kl})\nhat^k\nhat^l + \frac2\chi h'_{kl} m_\mp^k \nhat^l\right).
\ea
We then decompose $h_{ij}$ into Fourier modes of two polarization states
(see \cite{paperI,paperII} for details),
\ba
h_{ij}(\vk, \eta) =\:& e^+_{ij}(\hat\vk) h^+(\vk,\eta) + e^\times_{ij}(\hat\vk)
h^\times(\vk, \eta),
\label{eq:hpol}
\ea
where $e^s_{ij}(\hat\vk)$, $s=+,\times$, are transverse 
(with respect to $\hat\vk$) and 
traceless polarization tensors normalized through
$e^s_{ij} e^{s'\:ij} = 2 \d^{ss'}$.  We assume both polarizations to
be independent and to have equal power spectra:
\ba
\< h_{s}(\vk,\eta) h_{s'}(\vk',\eta') \> =\:& (2\pi)^3 \d_D(\vk-\vk') 
\d_{ss'} \frac14 T_T(k,\eta) T_T(k,\eta') P_{T0}(k),
\label{eq:PT}
\ea
where $T_T(k,\eta)$ is the tensor transfer function, and the primordial
tensor power spectrum is denoted as $P_{T0}(k)$.  Further, we can define
helicity$\pm 2$ polarization tensors and
Fourier amplitudes through
\ba
e^{\pm 1}_{ij} \equiv\:& e^+_{ij} \pm i e^\times_{ij} \vs
h_{\mp 1} \equiv\:& \frac12 (h^+ \pm i h^\times).
\ea
Note that $P_{h_{\pm 1}}(k) = P_{T0}(k)/8$.  
As before, we begin by evaluating the contribution of
a single plane wave, assuming that $\vk = k \hat{\v{z}}$.  We have
\ba
e^p_{\pm}(\hat\vk,\vnhat) \equiv e^p_{ij}(\hat\vk) m_\mp^i(\vnhat) m_\mp^j(\vnhat) =\:& \frac12 (1\mp p \mu)^2 e^{i2p\phi}
\vs
e^p_\parallel(\hat\vk) \equiv e^p_{ij} \nhat^i \nhat^j  =\:& 
(1-\mu^2) e^{i2p\phi}
\vs
e^p_{ij}(\hat\vk) m_\pm^i(\vnhat) \nhat^j =\:& 
\frac{\sqrt{1-\mu^2}}{\sqrt{2}} (\mu \pm p) e^{i2p\phi},
\ea
where $\mu=\cos\theta$.  
Restricting to the polarization $p=+1$ first, we have
\ba
{}_{\pm1}\B(\vk,\vnhat,+1) =\:&  
\frac{1+\zt}{2 H(\zt)}\int_0^{\chit}d\chi\frac{\chi}{\chit}\left\{
-\frac{\sqrt{1-\mu^2}}{\sqrt{2}} (1-\mu^2) i k
+ \frac{\sqrt2}{\chi} \sqrt{1-\mu^2} (\mu\mp 1)\right\} 
h'_1(\vk,\eta) e^{i2\phi} e^{i x\mu}.
\ea
We now apply the spin-lowering operator \refeq{Delmu} with $m=2$ to obtain
\ba
\bar\Del {}_1\B(\vk,\vnhat,+1) =\:& \left[\partial_\mu - \frac{2}{1-\mu^2}\right] \sqrt{1-\mu^2} {}_1 \B(\vk,\vnhat,+1) \vs
=\:& \left[\partial_\mu - \frac{2}{1-\mu^2}\right]\Bigg\{
\frac{1+\zt}{2 H(\zt)}\int_0^{\chit}\frac{d\chi}{\chit}\left\{
-\frac{(1-\mu^2)^2}{\sqrt{2}} i x
- \sqrt2 (1-\mu^2) (1-\mu)  \right\} 
h'_1(\vk,\eta) e^{i2\phi} e^{i x\mu}\Bigg\}
\vs
=\:& 
\frac{1+\zt}{2 H(\zt)}\int_0^{\chit}\frac{d\chi}{\chit}\left\{
\frac1{\sqrt2} \hat Q_{BT1}(x) + \sqrt2 \hat Q_{BT2}(x)
\right\} 
(1-\mu^2) e^{i2\phi} e^{i x\mu}\Bigg\} h'_1(\vk,\eta) \vs
\label{eq:Bktensor}
\ea
where
\ba
\hat Q_{BT1}(x) =\:& x^2 + 4 x \partial_x + x^2 \partial_x^2 + 2 i x
\vs
\hat Q_{BT2}(x) =\:& 3 + x \partial_x - ix.
\ea 
For the other polarization state, we obtain the same result 
with $\mu \to -\mu$, $x \to - x$, $\phi \to -\phi$.   
\refeq{Bktensor} is in the desired form, \refeq{step2}.  We see that
the operators have both real and imaginary parts, signaling that
tensor modes contribute to both the polar (``$E$-mode'', 
through $\Re \hat Q_{BTi}$) and axial vector (``$B$-mode'', through $\Im \hat Q_{BTi}$).  

\section{Components in terms of matter density contrast}
\label{app:calc}

This section gives useful expressions for $\C$, ${}_{\pm 1}\B$, $\M$,
and the shear in terms of the familiar matter density contrast $\d_m^{\rm sc}$ in 
synchronous-comoving gauge.  In the following, we will neglect terms that only contribute to monopole and dipole, since they are observationally irrelevant in essentially all cases.  
For convenience we write the velocity $v_i$ in terms of a
scalar velocity potential $V$, $v_i = V_{,i}$, and relate $V,\,\Phi,\,\Psi$
to the density contrast $\d_m^{\rm sc}$ in synchronous-comoving gauge
through (see \cite{Schmidt08,HuSaw07b})
\ba
V(\vk,\eta) =\:& a H f k^{-2} D(a(\eta)) \d_m^{\rm sc}(\vk,\eta_0)
\vs
\Phi(\vk,\eta) -\Psi(\vk,\eta) =\:& D_{\Phi_-}(k,\eta) \d_m^{\rm sc}(\vk,\eta_0) \vs
\Phi(\vk,\eta) + \Psi(\vk,\eta) =\:& g(k,\eta) D_{\Phi_-}(k,\eta) \d_m^{\rm sc}(\vk,\eta_0) \vs
\Rightarrow \Phi(\vk,\eta) =\:& \frac12 [g+1] D_{\Phi_-} \d_m^{\rm sc}(\vk,\eta_0) \vs
 \Psi(\vk,\eta) =\:& \frac12 [g-1] D_{\Phi_-} \d_m^{\rm sc}(\vk,\eta_0) \vs
(\Phi-\Psi)'(\vk,\eta) =\:& D_{\rm ISW}(k,\eta) \d_m^{\rm sc}(\vk,\eta_0),
\ea
where $f \equiv d\ln D/d\ln a$, $D(a)$ is the matter growth factor
(normalized to unity at $a=1$) and we have defined general coefficient
functions $D_{\Phi_-},\,g,\,D_{\rm ISW}$ to allow for non-standard 
cosmologies.  In a $\Lambda$CDM cosmology (or more generally for
a smooth dark energy component), we have
\ba
D_{\Phi_-}(k,\eta) =\:
&
3 \Omega_m \frac{a^2H^2}{k^2} D(a(\eta))
=
3 \Omega_{m0} \frac{H_0^2}{k^2} a^{-1}(\eta) D(a(\eta))\vs
g(\vk,\eta) =\:& 0 \vs
D_{\rm ISW}(k,\eta) =\:& \frac{\partial}{\partial\eta} D_{\Phi_-}(k,\eta)\,.
\ea
Here, a subscript $0$ denotes that the quantity is defined at the present
epoch $\eta = \eta_0$.  We will denote the power spectrum of $\d_m^{\rm sc}$ 
at $z=0$ as $P_m(k)$, as well as $x=k\chi$, $\tilde x = k\chit$ as before.    

\subsection{Longitudinal scalar}

We begin with \refeq{C_cN},
\ba
(\C)_{\rm cN} =\:& -b_z\,\D\ln a
 - \Psi - v_\parallel + \frac{1+\zt}{H(\zt)} \left(
 - \partial_\parallel\Psi + \partial_\parallel v_\parallel 
 - v_\parallel' + \Phi'  \right),
\ea
where
\ba
b_z =\:& 1-H(\zt)\partial_{\zt}\left(\frac{1+\zt}{H(\zt)}\right) \vs
\D\ln a =\:& \Psi_o - \Psi + v_\parallel - v_{\parallel o}
+ \int_0^{\chit} d\chi \left[ \Phi' - \Psi' \right]\,,
\ea
and we have dropped the monopole term $\propto \int \Psi(\v{0},t)dt$.  
Note that for a non-evolving ruler in $\Lambda$CDM, $b_z = 3\Omega_m(\zt)/2$.  
For a single plane-wave perturbation, this yields
\ba
\C(\vk,\vnhat) =\:
& -b_z \Bigg\{
\left[
\left(
\Psi(\vk,\eta_0)-i\mu x\frac{V(\vk,\eta_0)}{\chi}
\right)e^{i x \mu} \right]_{x=0}
- \left(
\Psi(\vk,\tilde\eta)-i\mu\tilde{x}\frac{V(\vk,\tilde\eta)}{\chit}
\right)e^{i\tilde x \mu}
+ \int_0^{\chit} d\chi \left[
\Phi' - \Psi'
\right]e^{ix\mu}
\Bigg\} 
\vs
& - \left[\Psi(\vk,\tilde\eta) + i \mu \tilde x \frac{V(\vk,\tilde\eta)}{\chit}\right]
e^{i\tilde x\mu}  + \frac{1+\zt}{H(\zt)\chit} \left[
- i \mu \tilde x \Psi(\vk,\tilde\eta) - \mu^2 \tilde x^2 \frac{V(\vk,\tilde\eta)}{\chit}
- i \mu \tilde x V'(\vk,\tilde\eta) + \chit\Phi'(\vk,\tilde\eta)\right] e^{i \tilde x \mu}
\vs
=\:& 
\delta_m^{\rm sc}(\vk,\eta_0) 
 (-b_z)
\Biggl\{
\left[
\left(
\frac12(g-1)D_{\Phi_-}
-\frac{aHfD}{k^2\chit}\hat Q_{CS1}(\tilde{x})
\right)
e^{i x \mu}
\right]_{x=0} 
\vs 
&
\hspace{2.4cm}
- \left(
\frac12(g-1)D_{\Phi_-}
-\frac{aHfD}{k^2\chit}\hat Q_{CS1}(\tilde{x})
\right)
+ \int_0^{\chit} d\chi D_{\rm ISW}e^{ix\mu} \Biggl\}
\vs
&+
\d_m^{\rm sc}(\vk,\eta_0)
\Bigg\{-\left[\frac12 (g-1) D_{\Phi_-} + \frac{aHf D}{k^2 \chit} \hat Q_{CS1}(\tilde x) \right] \vs
& \hspace*{2.2cm} + \frac{1+\zt}{H(\zt)\chit} \bigg[
- \frac12 (g-1) D_{\Phi_-} \hat Q_{CS1}(\tilde x) +  \frac{a H f D}{k^2 \chit} \hat Q_{CS2}(\tilde x)
-  (a H f D)' \hat Q_{CS1}(\tilde x) \vs
& \hspace*{3.9cm} + \frac{\chit}2 \left(g'D_{\Phi_-} + (g+1)D_{\rm ISW}\right) \bigg]\Bigg\}_{k,\tilde\eta} e^{i \tilde x \mu},
\ea
where $\hat Q_{CS1}(x) = x \partial_x$, and $\hat Q_{CS2}(x) = x^2 \partial_x^2$. 
This is clearly in the form \refeq{step2}, with $r=s=0$,  and we can thus
immediately apply \refeq{CAXX}:
\ba
C_{\C}(l) =\:& \frac2\pi \int k^2 dk P_m(k) |F_l^{\C}(k)|^2 \vs
F_l^{\C}(k) =\:& - b_z F_l^{\D\ln a}(k)
+
\Bigg\{-\left[\frac12 (g-1) D_{\Phi_-} + \frac{aHf D}{k^2 \chit} \hat Q_{CS1}(\tilde x) \right] \vs
& \hspace*{2.5cm} + \frac{1+\zt}{H(\zt)\chit} \bigg[
- \frac12 (g-1) D_{\Phi_-} \hat Q_{CS1}(\tilde x) +  \frac{a H f D}{k^2 \chit} \hat Q_{CS2}(\tilde x)
-  (a H f D)' \hat Q_{CS1}(\tilde x) \vs
& \hspace*{4.2cm} + \frac{\chit}2 \left(g'D_{\Phi_-} + (g+1)D_{\rm ISW}\right) \bigg]\Bigg\}_{k,\tilde\eta}
j_l(\tilde x)
\vs
F_l^{\D\ln a}(k) \equiv\:& \left(\frac{a H f D}{k}\partial_{\tilde x} - \frac12(g-1) D_{\Phi_-} \right)_{\zt} j_l(\tilde x)
 + \int_0^{\chit} d\chi D_{\rm ISW} j_l(x)
\,.
\label{eq:Fdlna}
\ea

\subsection{Vector}

As derived in \refapp{vector},
\ba
\hspace*{-0.2cm}\bar\Del {}_1\B(\vnhat,\vk) =\:&  
\frac{V(\vk,\tilde\eta)}{\sqrt2 \chit}
\hat Q_{BS1}(\tilde x) e^{i \tilde x\mu}
 + \frac{1+\zt}{\sqrt2 H(\zt)}\Bigg\{
\frac1{\chit} \Psi(\vk,\tilde\eta) \hat Q_{BS1}(\tilde x) 
 e^{i\tilde x \mu}
+ k^2 V(\vk,\tilde\eta) \left[\hat Q_{BS2}(\tilde x) - \frac1{\tilde x^2}\hat Q_{BS1}(\tilde x) \right] e^{i\tilde x \mu}
\vs
& \hspace*{4.5cm}
 + \frac1{\chit^2} \left[\hat Q_{BS1}(x) e^{i x\mu}\right]_
{x=0}\!\!\!\!\! V(\vk,\eta_0)
- \int_0^{\chit} \frac{d\chi}{\chit} \hat Q_{BS1}(x) [\Phi'(\vk,\eta)-\Psi'(\vk,\eta)]e^{i x \mu}
\Bigg\} \vs
 =\:&  \d_m^{\rm sc}(\vk,\eta_0) \Bigg\{ \bigg[
\frac{a H f D}{\sqrt2 k^2\chit} \hat Q_{BS1}(\tilde x) \vs
& \hspace*{1.9cm} 
+ \frac{1+\zt}{\sqrt2 H(\zt)}\left(\frac{1}{2\chit} [g-1] D_{\Phi_-}(k,\tilde a) \hat Q_{BS1}(\tilde x) 
+ a H f D\left[\hat Q_{BS2}(\tilde x)-\frac1{\tilde x^2}\hat Q_{BS1}(\tilde x)\right]\right) \bigg]_{\tilde\eta}\!\! e^{i\tilde x \mu}
\vs
& \hspace*{1.7cm} 
+ \frac{1+\zt}{\sqrt2 H(\zt)}\left[ 
(H f D)_{z=0} \frac1{\tilde x^2} \left[\hat Q_{BS1}(x) e^{i x\mu} \right]_{x=0}
-  \int_0^{\chit} \frac{d\chi}{\chit}    
D_{\rm ISW}(k, a(\chi))\, \hat Q_{BS1}(x) e^{i x \mu} \right]
 \Bigg\}. 
\ea
Since ${}_{\pm 1}\B$ is a spin-1 quantity, \refeq{CAXX} with $r=0,\,s=1$ yields
\ba
C^{EE}_{\B}(l) =\:& \frac2\pi \frac1{l(l+1)}
\int k^2 dk P_m(k) |F_l^{\B E}(k)|^2 \vs
F_l^{\B E}(k) =\:& 
\bigg[
\frac{a H f D}{\sqrt2 k^2\chit} \hat Q_{BS1}(\tilde x) 
+ \frac{1+\zt}{\sqrt2 H(\zt)}\left(\frac{1}{2\chit} [g-1] D_{\Phi_-}(k,\tilde a) \hat Q_{BS1}(\tilde x) 
+ a H f D\left[\hat Q_{BS2}(\tilde x)-\frac1{\tilde x^2}\hat Q_{BS1}(\tilde x)\right]\right) \bigg]_{\tilde\eta} j_l(\tilde x)
\vs
& + \frac{1+\zt}{\sqrt2 H(\zt)}\left[ 
(H f D)_{z=0} \frac1{\tilde x^2} \left[\hat Q_{BS1}(x) j_l(x) \right]_{x=0}
-  \int_0^{\chit} \frac{d\chi}{\chit}    
D_{\rm ISW}(k, a(\chi))\, \hat Q_{BS1}(x) j_l(x) \right]. 
\ea
This is the power spectrum of the $E$-mode (polar) component of 
${}_\pm \B$.  The operators $\hat Q_{BSi}$ as applied to spherical
Bessel functions become
\ba
\hat Q_{BS1}(x) j_l(x) =\:& -l(l+1) j_l(x) \vs
\hat Q_{BS2}(x) j_l(x) =\:& 
l(l+1) \partial_x\left(\frac{j_l(x)}{x}\right).
\ea
Note that this implies $[\hat Q_{BS2}(x) - x^{-2} \hat Q_{BS1}(x)] j_l(x) = l(l+1) (\partial_x j_l(x))/x$.  Note further that the observer term
$[\hat Q_{BS1}(x) j_l(x)]_{x=0}$ is only non-zero for the dipole $l=1$, as
expected.  

\subsection{Shear and magnification}

We begin with the shear.  
\refeq{shear_cN} evaluated for a plane-wave perturbation yields
\ba
{}_{\pm 2} \g(\vnhat,\vk) =\:& 
- \int_0^{\chit} d\chi\, (\chit-\chi)\frac{\chi}{\chit}
 \frac12 (1-\mu^2) k^2 \left[\Psi - \Phi \right]_{\vk,\eta} e^{ix\mu}
\vs
=\:& \frac12 \int_0^{\chit} d\chi\, (\chit-\chit)\frac{\chi}{\chit} k^2
 D_{\Phi_-}(k,a(\chi)) (1-\mu^2) e^{ix\mu}\: \d_m^{\rm sc}(\vk,\eta_0).
\ea
In order to obtain a scalar quantity, we now apply \refeq{Del} with $m=0$, 
yielding
\ba
\bar\Del^2 {}_{2} \g(\vnhat,\vk) =\:&
\partial_\mu^2 [ (1-\mu^2) {}_2 \g] \vs
=\:& \frac12 \int_0^{\chit} d\chi\, (\chit-\chi)\frac{\chi}{\chit}
 k^2 D_{\Phi_-}(k,a(\chi)) 
\partial_\mu^2\left[ (1-\mu^2)^2 e^{ix\mu}\right]\: \d_m^{\rm sc}(\vk,\eta_0) \vs
=\:& \frac12 \int_0^{\chit} d\chi\, (\chit-\chi)\frac{\chi}{\chit}
 k^2 D_{\Phi_-}(k,a(\chi)) \hat Q_{\g S}(x) e^{i x\mu}
\: \d_m^{\rm sc}(\vk,\eta_0),
\ea
where we have defined
\be
\hat Q_{\g S}(x) = 4 + x^2 + 8 x \partial_x + (12+2x^2) \partial_x^2
+ 8x\partial_x^3 + x^2 \partial_x^4 .
\ee
This is again in the desired form \refeq{step2}, and applying \refeq{CAXX}
with $r=0,\,s=2$ yields the angular power spectrum of the shear $E$-modes,
\ba
C^{EE}_{\g}(l) =\:& \frac2\pi \frac{(l-2)!}{(l+2)!} 
\int k^2 dk P_m(k) |F_l^{\g E}(k)|^2 \vs
F_l^{\g E}(k) =\:& \frac12 \int_0^{\chit} d\chi\, (\chit-\chi)\frac{\chi}{\chit}
 k^2 D_{\Phi_-}(k,a(\chi)) \hat Q_{\g S}(x) j_l(x).
\ea
The operator $\hat Q_{\g S}$ applied to spherical Bessel functions is
\ba
\hat Q_{\g S}(x) j_l(x) = \frac{(l+2)!}{(l-2)!} \frac{j_l(x)}{x^2}.
\ea

The magnification contains a number of terms, which are evaluated in
Appendix~B of \cite{paperI}.  Here we only repeat the final result:
\ba
C_{\M}(l) =\:& \frac2\pi \int k^2 dk\,P_m(k) |F_l^{\M}(k)|^2 
\vs
F_l^{\M}(k) =\:& 2\left[\frac1{a H \chit} - 1\right] F_l^{\D\ln a}(k) 
- ([g+1] D_{\Phi_-})_{\zt}\, j_l(\tilde x) \vs
& + l(l+1) \int_0^{\chit} d\chi \frac{\chit-\chi}{\chi\chit} D_{\Phi_-} j_l(x) 
+ 2 \int_0^{\chit}\frac{d\chi}{\chit} D_{\Phi_-} j_l(x)
\label{eq:Clmag},
\ea
where $F_l^{\D\ln a}(k)$ is defined in \refeq{Fdlna}.

\end{widetext}
\bibliography{GW}

\end{document}